\documentclass[aps,twocolumn,showlabels,showrefs,amsmath,amssymb,prl,superscriptaddress,floatfix,colors]{revtex4-2}

\usepackage{lineno}
\usepackage{graphicx}
\usepackage{dcolumn}
\usepackage{bm}
\usepackage{cancel}
\usepackage{graphicx}

\usepackage{dcolumn}
\usepackage{bm}
\usepackage{amssymb}
\usepackage{hyperref}

\usepackage{multirow}
\usepackage{color}
\usepackage[normalem]{ulem}
\newcommand{\ylann}[1]{{\color{black} #1}}

\usepackage[cp1251]{inputenc}

\makeatletter
\newcommand*{\sumcirclearrowleft}{%
 \DOTSB
 \mathop{
  \mathchoice
   {\rlap{\kern.25em\rotatebox[origin=c]{-90}{$\circlearrowleft$}}{\sum}}
   {\vcenter{\rlap{\kern.2em\rotatebox[origin=c]{-90}{$\scriptscriptstyle\circlearrowleft$}}}{\sum}}
   {\sum}{\sum}
 }\slimits@
}

\newcommand*{\sumcirclearrowright}{%
 \DOTSB
 \mathop{
  \mathchoice
   {\rlap{\kern.25em\rotatebox[origin=c]{90}{$\circlearrowright$}}{\sum}}
   {\vcenter{\rlap{\kern.2em\rotatebox[origin=c]{90}{$\scriptscriptstyle\circlearrowright$}}}{\sum}}
   {\sum}{\sum}
 }\slimits@
}
\makeatother

\begin{document}

\title{Dynamics of $O(2)$ excitations in a  non-reciprocal medium}


\author{Ylann Rouzaire} \email{Corresponding author: rouzaire.ylann@gmail.com}
 \affiliation{Departament de F\'isica de la Materia Condensada, Universitat de Barcelona, Mart\'i i Franqu\`es 1, E08028 Barcelona, Spain}
  \affiliation{UBICS University of Barcelona Institute of Complex Systems , Mart\'i i Franqu\`es 1, E08028 Barcelona, Spain}
   \affiliation{Computing and Understanding Collective Action (CUCA) Lab, Mart\'i i Franqu\`es 1, E08028 Barcelona, Spain}%

 \author{Daniel J.G. Pearce}
 \affiliation{Universit\'e de Gen\`eve D\'{e}partement de physique th\'{e}orique, 24 quai Ernest-Ansermet, 1211 Gen\`{e}ve, Switzerland }
 
 \author{Ignacio Pagonabarraga} 
 \affiliation{Departament de F\'isica de la Materia Condensada, Universitat de Barcelona, Mart\'i i Franqu\`es 1, E08028 Barcelona, Spain}
  \affiliation{UBICS University of Barcelona Institute of Complex Systems , Mart\'i i Franqu\`es 1, E08028 Barcelona, Spain}

 \author{Demian Levis} 
 \affiliation{Departament de F\'isica de la Materia Condensada, Universitat de Barcelona, Mart\'i i Franqu\`es 1, E08028 Barcelona, Spain}
 \affiliation{UBICS University of Barcelona Institute of Complex Systems , Mart\'i i Franqu\`es 1, E08028 Barcelona, Spain}
 \affiliation{Computing and Understanding Collective Action (CUCA) Lab, Mart\'i i Franqu\`es 1, E08028 Barcelona, Spain}%

\date{\today}
\begin{abstract}
We investigate emergent dynamics due to non-reciprocity in the $\mathcal{O}(2)$ model. The lattice XY model, where non-reciprocity stems from vision cone like couplings, can be described by a continuum description in which non-reciprocity translates into a new term depending on the rotational of the orientation field. We argue that non-reciprocity is akin to activity and we highlight the connection between our hydrodynamic equation and the constant density Toner-Tu framework.
The active force advects and reshapes patterns, a generic feature found in many non-reciprocal systems. 
We show how $1d$ excitations in the non-reciprocal $\mathcal{O}(2)$ model can be  described by a generalized Burgers equation,  derived from our continuum model. We then extend the results to $2d$ perturbations. As such, we establish the first principles of excitation trajectory control in a non-reciprocal $\mathcal{O}(2)$ medium. Concretely, we explain how tuning the degree of non-reciprocity and the orientation of the background medium impacts the time evolution  of excitations. We also showcase how initially different excitations lead to very different dynamical behavior. 
Non-reciprocity also affects the stability of defect-free excitations with non-zero winding numbers and, unlike in its equilibrium $O(2)$ counterpart, enables the system, above a certain threshold, to relax to its ground state.
\end{abstract}
\maketitle

\vspace{1cm}

Effective non-reciprocal interactions between coarse-grained objects usually arise in out-of-equilibrium conditions. They can arise in many-species systems, where the goals of the agents and the interactions between them are not necessarily symmetric \cite{saha2020nonreciprocalcahnhilliard,  you2020nonreciprocity, fruchart2021non,  dinelli2023non, mandal2024robustness, kreienkamp2022clustering, mason2025dynamical, alvarez2025segregation}. Another route towards non-reciprocity can be found in single species populations in which agents perceive their surroundings through an anisotropic field of view. This is indeed a common feature in animal groups and typically impacts the collective motion \cite{pita2015vision}. However, in those systems, agents typically move in space, and the role played by non-reciprocity and self-propulsion are intimately entangled. 
The study of self-propelled units in the framework of physics dates back to 1995, when Vicsek and coauthors introduced a minimal agent-based model \cite{vicsek1995novel}  exhibiting spontaneous collective motion, or flocking. Later that year, Toner and Tu \cite{toner1995long} proposed a continuum theory to  understand the large scale properties of such flocks, in particular in two spatial dimensions.   
More recently, a growing literature has focused on non-reciprocal, yet immobile, single-species models \cite{dadhichi_nonmutual_2020,loos2023long, rouzaire2025nonreciprocal, popli2025ordering, bandini2024xy,garces2025phase, agarwal2026higher}, as well as associated coarse-grained descriptions \cite{vafa2022defect, dadhichi_nonmutual_2020, besse2022metastability,  dopierala2025inescapable, huang2024active, agarwal2026higher}.
\ylann{
These works largely focussed on the nature of the ordered phase in the steady-state~\cite{loos2023long, bandini2024xy, popli2025ordering}, its stability to fluctuations \cite{dadhichi_nonmutual_2020, besse2022metastability, popli2025ordering} and its anisotropy at large scales in lattice based models \cite{bandini2024xy, popli2025ordering, dopierala2025inescapable}. 

Topological defects constitute the second axis of the literature on single species non-reciprocity. In addition to their role in the destabilisation of the order, they are also natural objects to study in that context, as their collective behaviour triggers the BKT topological phase transition in the equilibrium limit (the reciprocal $2d$ XY model). In contrast with the shape-agnostic dynamics of the XY defects, under the action of non-reciprocity $+1$ defects decay to specific shapes \cite{vafa2022defect, rouzaire2025nonreciprocal, popli2025ordering} and $-1$ defects develop extended axes out from their core \cite{besse2022metastability, rouzaire2025nonreciprocal}. Combined, these dynamics lead to new defect annihilation scenarios, informing on the breakdown of the BKT phase transition mechanisms and coarsening dynamics \cite{rouzaire2025nonreciprocal}. 

Beyond the study of defects, our understanding of the dynamics in such systems is relatively scarce, despite a rich phenomenology. In particular, non-reciprocal systems generically feature unidirectional pattern propagation. 
A steady propagative state in a dissipative medium is only possible if the system obeys conservation laws. This is the case of multi-species particle mixtures, for which a non-reciprocal extension of the Cahn-Hilliard model  has been proposed as a  hydrodynamic description, and has been shown to exhibit traveling bands in the steady-state, akin to the chase-and-run long-term dynamics in prey-predator like models  \cite{you2020nonreciprocity, saha2020nonreciprocalcahnhilliard, chiu2023phase, mandal2024robustness, weis2025generalizedNRmultipopulation, pisegna2024emergent}.
In absence of a conserved field (in the latter case, the density of each species), propagative states are formally only transient. Yet, even in this case, perturbation can travel across large distances before vanishing \cite{loos2023long}, motivating a further dedicated study on their dynamics.  

This ubiquitous feature is a direct consequence of the (self-)advection term that naturally arises from the microscopic dynamics.
This active term appears in many different models and does not depend on the details of the source of non-reciprocity.
It arises in spin models where the interaction kernel, independent of the state of the spin, is spatially translated \cite{seara2023NRIsing}. 
It also follows from vision-cone \textit{social}-like interactions, for Ising spins \cite{garces2025phase, rajeev2024ising}, XY spins \cite{loos2023long, rouzaire2025nonreciprocal, bandini2024xy, popli2025ordering, dadhichi_nonmutual_2020} or non-reciprocal \textit{steric} interactions \cite{huang2024active}.  
In non-polar mixtures with non-reciprocal interactions, the emergent polar order is also self-advected in the same fashion  \cite{pisegna2024emergent}.
Finally, we conjecture that a similar self-advection term is also at the root of the flows recently reported in \cite{klamser2025directed}. 
\newline

To shed light on this phenomenology, we analyse a non-reciprocal extension of the $O(2)$ model, as it allows both efficient numerical simulations and tractable analytical treatments. 
Within this framework, we investigate the impact of non-reciprocity on the dynamics of smooth excitations, or non-uniform profiles. We neglect thermal fluctuations to isolate the core impact of non-reciprocity.
}   
The manuscript is organized as follows. 
To provide a microscopic motivation of the model, we present in section 1 the non-reciprocal $2d$XY (lattice) model and how one can derive a continuum description from it. We discuss the  nature of the model in the context of field theories of polar active matter. 
In section 2, we describe how localized perturbations evolve in our non-reciprocal $2d$  system. We provide a full characterization of their dynamics: both  how they travel while decaying, as well as  why they generically develop asymmetric shapes. We first focus on $1d$ perturbations before extending our results to $2d$ ones.
In section 3, we describe the impact non-reciprocity has on spinwave configurations. We explain how a sufficiently high non-reciprocity destroys those excitations, allowing the system to relax to its equilibrium ground state.

\section{1. The model}
\begin{figure}[b!]
    \centering
    \includegraphics[width=\linewidth]{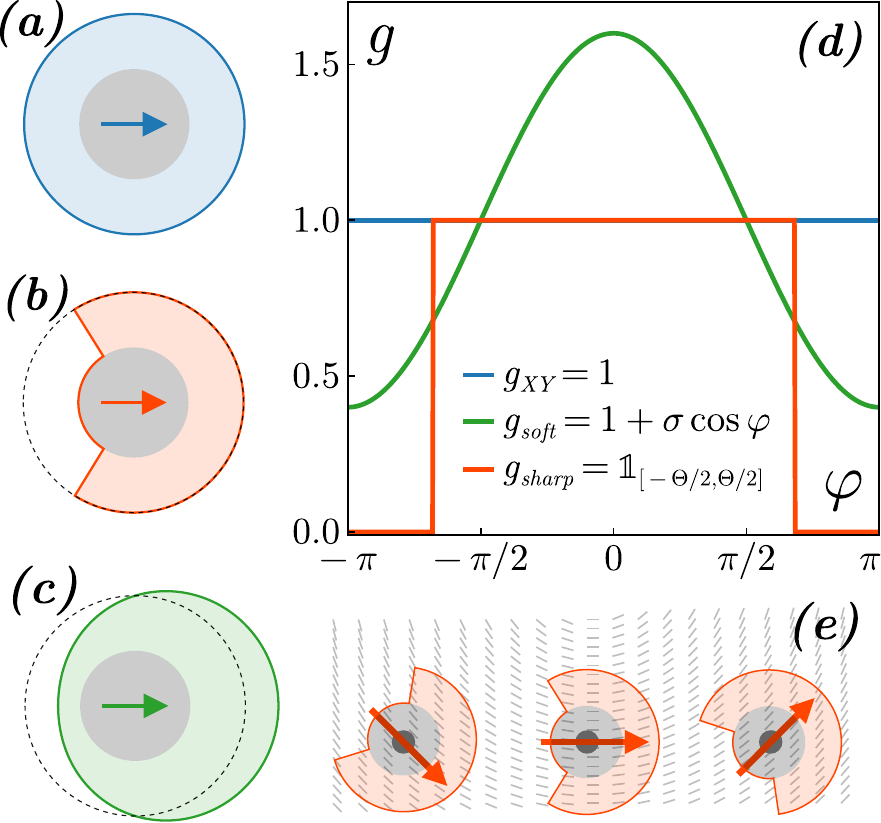}
    \caption{Three possible kernels describing the angular dependence of the coupling strength: \textbf{(a)} a uniform kernel \textbf{(b)} a sharp vision cone, taking value 1 in front and 0 behind. Here, the vision cone aperture is $\Theta=2\pi-2$ . \textbf{(c)} a smooth vision cone, smoothly changing from $1+\sigma$ (in front) to $1-\sigma$ (behind), here with $\sigma = 0.6$.  All the  kernels we study are centered on the current orientation of the spin, here depicted by the horizontal arrows. The dotted lines in panels (b, c) represent the uniform kernel in (a). \textbf{(d)} The three kernels illustrated in (a)-(c) as functions of $\varphi_{ij}=\text{Angle}(\boldsymbol{\hat S}_i\cdot  \boldsymbol{u}_{ij})$.  
     \textbf{(e)} Illustration of the impact of a positive rotational vector field on an individual spin; see main text for discussion.}
    \label{fig:kernels}
\end{figure}

This work focuses on a continuous field theory, conceived as a non-reciprocal extension of the classical $O(2)$ model \cite{mussardo2010statistical}. To provide a microscopic motivation of the latter, we start by considering a two-dimensional (2$d$) XY model with vision cone interactions \cite{rouzaire2025nonreciprocal}. 
The system  is composed of spins $\boldsymbol{\hat S}_i=(\cos\theta_i,\sin\theta_i)$ sitting on the nodes of a regular lattice of linear size $N$. 
Two neighboring spins will tend to align their respective orientation in a ferromagnetic way, such that their phase $\theta$ evolves according to 
\begin{equation}
\gamma\,\dot{\theta}_i=J\sum_{j \in \partial_i} \,g(\varphi_{i j})\sin \left(\theta_j-\theta_i\right) 
 \label{eq:eom}
 \end{equation}
 where $\gamma$ is the damping coefficient and $J$ the coupling strength. The kernel $g$ encodes the anisotropic nature of the interactions and depends on the angle $\varphi_{ij}$ between the orientation of the spin $i$ and the director of the ($ij$)-bond $\boldsymbol{u}_{ij} = \boldsymbol{r}_{j} - \boldsymbol{r}_{i}$. Thus $\varphi_{ij} = \text{Angle}(\boldsymbol{\hat S}_i\, ,   \boldsymbol{u}_{ij})$. 
The kernel $g$ can be chosen freely, depending on the problem at hand. We sketch in Fig.~\ref{fig:kernels}(a-c) three possible kernels centered on the orientation of the spin.

In panel Fig.~\ref{fig:kernels}(a), the kernel is constant $g(\varphi) = 1 $ and the model boils down to the equilibrium $O(2)$ model, the continuum version of the $2d$XY model. 
To describe our model at coarse-grained scales, 
we first introduce the vector field $\boldsymbol{S}(x, y, t)=S(\cos\theta,\sin\theta)$. Now,  $\theta=\theta(x,y, t)$ is  the local phase of the field and  $S=|\boldsymbol{S}(x,y, t)|$ its modulus. The scalar order parameter $S$ corresponds to the local polarization averaged over some coarse-graining length scale.
Following  \cite{YurkeHuse1993}, the dynamics of $\boldsymbol{S}$ in the equilibrium $O(2)$ model are governed by 
\begin{equation}
    \gamma\dot{\boldsymbol{S}} = K\Delta\boldsymbol{S} + \alpha (1-\boldsymbol{S}^2)\boldsymbol{S} \ ,
    \label{eq:continuousXY} 
\end{equation}
where $\gamma$ is the damping coefficient and $K/\gamma$ the diffusion constant.
Equation~(\ref{eq:continuousXY}) can be written as a relaxational dynamics in the presence of dissipation with the following Ginzburg-Landau free energy functional \cite{YurkeHuse1993} 
\begin{equation}
    \mathcal{F}[\boldsymbol{S}] = \int d\textbf{r} \left[ \frac{K}{2} (\nabla\boldsymbol{S})^2 +\frac{\alpha}{4}\left(1-\boldsymbol{S}^2\right)^2 \right]\, .
    \label{eq:free_energy}
\end{equation}
The first term in Eq.~(\ref{eq:free_energy}) penalizes gradients in the director field $\boldsymbol{S}$, pushing the system towards uniform  states. The second term penalizes deviations of the scalar order parameter from unity. 
 The parameter $\alpha$ is thus the Lagrange multiplier associated to the spherical constraint and gives rise to the length scale $\lambda_\alpha = \sqrt{K/\alpha}$ over which  $S$ can smoothly vary from $0$ to $1$ to balance the excess free energy associated to non-uniform profiles. The $2d$XY limit corresponds to the strict spherical constraint, $S^2=1$, i.e.  $\alpha\to \infty$ and $\lambda_\alpha\to 0$. 

This equilibrium XY scenario is greatly enriched by the introduction of anisotropic couplings, through the non-reciprocal kernel $g$. 
In panel (b), the kernel is given by a step function $g(\varphi; \Theta) = 1 \text{ if } |\varphi| \le \Theta/2  $ and represents a sharp vision cone. An aperture $\Theta < 2\pi$ creates a blind spot at the rear and makes the interactions generically non-reciprocal, since a configuration where a spin $i$ reacts to another spin $j$ but $j$ does not react to $i$ violates the action-reaction principle between $i$ and $j$. This type of vision cone has been used in non-reciprocal agent-based models in \cite{loos2023long, bandini2024xy}, and to construct a hydrodynamic theory in \cite{huang2024active}.  

Panel (c) shows a smooth version of the vision cone $g(\varphi; \sigma) = 1 + \sigma \cos \varphi$, a kernel previously studied in~\cite{dadhichi_nonmutual_2020, rouzaire2025nonreciprocal, popli2025ordering, dopierala2025inescapable}. The parameter $\sigma$ controls the degree of non-reciprocity. Action-reaction is  recovered only for $\sigma~=~0$. 
\ylann{The explicit coarse-graining of this microscopic model leads to a hydrodynamic theory with undesired anisotropic terms \cite{popli2025ordering, dopierala2025inescapable}. 
Ref. \cite{rouzaire2025nonreciprocal} provided a technically simple, yet instructive, coarse-graining of the microscopic model on a square lattice based on Taylor expansions and trigonometric identities. This helped identify the main contribution of vision-cone non-reciprocity at larger scales. Numerical simulations confirmed that isolating this term indeed (quantitatively) reproduced the same physics. Therefore, and without pretending it to be a formal derivation from the microscopic model, this approach justifies the following hydrodynamic field theory: 
} 
\begin{equation}
\begin{aligned}
\gamma\dot{\boldsymbol{S}}&= - \frac{\delta \mathcal{F}}{\delta \boldsymbol{S}} + \bar{\sigma} (\nabla \times \boldsymbol{S})\times \boldsymbol{S} \\
&= K\Delta\boldsymbol{S}
\!+ \bar{\sigma} (\nabla \times \boldsymbol{S})\times \boldsymbol{S} + \alpha (1-|\boldsymbol{S}|^2)\boldsymbol{S} \ .
\end{aligned}
\label{eq:main_eq}
\end{equation}
{Here $\bar{\sigma}$ represents the general coarse-grained non-reciprocity, hence, such a non-reciprocal (NR) $\mathcal{O}(2)$ model is one in which non-reciprocity does not fade upon coarse-graining~\cite{dinelli2023non}. Without loss of generality, we focus on $\bar{\sigma} \ge 0$ since Eq.~(\ref{eq:main_eq}) {satisfies the same symmetries as Eq.~(\ref{eq:eom}) and is thereby } invariant under the joint transformation $\{\bar \sigma\to-\bar{\sigma}\ , \ \theta \to \theta+\pi\}$.}  
In the microscopic model, $|\sigma |>1$ introduces anti-alignment in the local interactions with neighbors at the back ($g(\pi,\sigma>1)<0$). To provide a sense of scale, the kernel $g(\varphi; \sigma) = 1 + \sigma \cos \varphi$ becomes negative in an angular region of aperture $2\pi-2\arccos(-1/\sigma)$. Thus a value of $\sigma =1.15$ implies 16\% of all interactions are antiferromagnetic (implying 1-2 neighbors on a triangular lattice), and this percentage climbs up to 33\% (2-3 neighbors) for $\sigma= 2$. This effect is not captured by Eq.~(\ref{eq:main_eq}), which assumes small variations between local spins. 

\subsection{Non-reciprocity is activity} 
Equation~(\ref{eq:main_eq}) highlights that non-reciprocity ultimately amounts to an active, state-dependent force, controlled by $\bar{\sigma}$. 
Indeed, the missing contribution of a neighbor $j$ in the blind spot of spin $i$ can be viewed as a virtual force canceling the force of $j$ on $i$ existing in equilibrium. We illustrate the dependence of the active term on the vorticity of the orientation field in Fig.~\ref{fig:kernels}(e), for three spins with vision cones in a field with positive rotational. In the equilibrium XY model, the middle spin would remain still because its two neighbors exert equal and opposite torques, reflecting the rotational invariance of the Laplacian operator.
Vision cones break this symmetry: removing the rear neighbor's contribution causes the spin to respond to the vorticity of the orientation field, explaining why a curl term arises when coarse-graining the dynamics. The odd derivative of the vector product translates the violation of the front/rear symmetry, in contrast with the even derivatives of the equilibrium Laplacian. 
This rotational force is an active, non-equilibrium  term, in that it cannot be derived from the gradient of a free energy. 

Beyond their activity, it is known (see \cite{dadhichi_nonmutual_2020, besse2022metastability}) that non-reciprocal XY models are essentially equivalent to the Toner-Tu equation for constant-density flocks \cite{toner2012birth, toner1995long}
\begin{equation}\label{eq:MtTT}
   \gamma\dot{\boldsymbol{S}}= - \frac{\delta \mathcal{F}}{\delta \boldsymbol{S}}   + \lambda_1(\boldsymbol{S}  \cdot \nabla) \boldsymbol{S} + \lambda_2( \nabla \cdot \boldsymbol{S}) \boldsymbol{S} + \lambda_3\nabla|\boldsymbol{S} |^2   \ .
\end{equation}
In the Toner-Tu framework, $\boldsymbol{S}$ represents the particles' velocity field. It is thus natural to find $\boldsymbol{S}$ involved as an advecting velocity in the $\lambda_1$ auto-advection term: particles advect information {(in the present case, the local value of $\boldsymbol{S}$)} through their motion.  In NRXY models however, spins do not move. Yet, the formal identity
\begin{equation}
 (\nabla \times \boldsymbol{S})\times \boldsymbol{S}=      (\boldsymbol{S}  \cdot \nabla) \boldsymbol{S} - \frac{1}{2}\,\nabla|\boldsymbol{S} |^2  \ 
 \label{eq:curl_adv}
\end{equation}
indicates that the non-reciprocity ultimately amounts to self-advection as well. This is a consequence of the kernel anisotropy, transporting a net information flux at the particle scale, from forward to rear, in the direction opposite to $\boldsymbol{S}$. A more careful coarse-graining based on Ito calculus \ylann{gives the linear relation between $\lambda_{1,2,3}$ and $\bar{\sigma}$}, see \cite{dopierala2025inescapable} and the Supplemental Material (SM)~\cite{SM}, section A. This leads to the identity $\lambda_3 = - \lambda_2 -\lambda_1 /2 $ and one can write Eq.~(\ref{eq:MtTT})  as 
\begin{equation}
   \gamma\dot{\boldsymbol{S}}= - \frac{\delta \mathcal{F}}{\delta \boldsymbol{S}}   + \lambda_1 (\nabla \times \boldsymbol{S})\times \boldsymbol{S} + \lambda_2\left(( \nabla \cdot \boldsymbol{S}) \boldsymbol{S} - \nabla|\boldsymbol{S} |^2  \right) \, . 
   \label{eq:chate_continuum}
\end{equation}
Finally, Eq.~(\ref{eq:chate_continuum}) can be re-written in the form
\begin{equation}
\begin{aligned}
       \gamma\dot{\boldsymbol{S}}= &\ - \frac{\delta}{\delta \boldsymbol{S}}\left[\mathcal{F} - \frac{\lambda_2 }{2}(\nabla\cdot\boldsymbol{S})|\boldsymbol{S}|^2\right] \\ 
       &+  \lambda_1 (\nabla\times\boldsymbol{S})\times\boldsymbol{S}
     - \frac{\lambda_2 }{2}\nabla|\boldsymbol{S}|^2 \ ,
\end{aligned}
\label{eq:withextraF}
\end{equation}
indicating that part of the $\lambda_2$ term can be written as the gradient of an additional free energy $\propto (\nabla \cdot\boldsymbol{S}) |\boldsymbol{S}|^2$ [see SM~\cite{SM}, section A for details on Eqs.~(\ref{eq:chate_continuum}) and (\ref{eq:withextraF})]. Writing the equation in this way highlights a second active effect. The $\lambda_2$ term locally aligns $\boldsymbol{S}$ with gradients in $\boldsymbol{S}^2$, thus has a large effect around topological defects. Indeed, the $\lambda_2<0$ found in \cite{dopierala2025inescapable} further enhances the stabilization of sink defects (inward asters), which are known to be the only stable $+1$ defects and the ones responsible for the slowdown of the annihilation dynamics  \cite{vafa2022defect, besse2022metastability, rouzaire2025nonreciprocal, popli2025ordering}. 

Setting $\lambda_2=0$ leaves us with Eq.~(\ref{eq:main_eq}) and allows us to focus on the effect of the $\lambda_1$ active term,  which we argue captures the main essence of non-reciprocity in our framework. First, we have shown in \cite{rouzaire2025nonreciprocal} that Eq.~(\ref{eq:main_eq}) reproduces the phenomenology of the lattice NRXY model Eq.~(\ref{eq:eom}), up to a quantitative agreement for the defect annihilation dynamics. 
Second, we focus here on continuous perturbations in the orientation of the vector field, which are associated with smaller variations in $S^2$.  
Third, the rotational term not only accounts for non-reciprocity (NR) stemming from \textit{reception cones} (eg. restricted vision), but also from \textit{emission cones} (eg. directed sound emission). 
Both microscopic models are different, yet they lead to the same hydrodynamic equation~(\ref{eq:main_eq}), \ylann{up to the transformation $\sigma \to -\sigma$}, see details in SM~\cite{SM} section B. In both cases, information is advected by the local polarization field.
This observation may prove useful in cases where it is experimentally simpler to design a setup reliant on directional emission rather than reception.

\section{2. Dynamics of smooth and localized excitations} \label{sec:propagation}

We now study the behavior of localized perturbations in an otherwise fully ordered system by simulating Eq.~(\ref{eq:main_eq}). We approximate the gradients using finite difference methods on a $256\times256$ periodic grid and integrate Eq.~(\ref{eq:main_eq}) using an Euler scheme with timestep $dt = 10^{-5}$. Unless otherwise stated, we set $\alpha = 100$, $\gamma=1$, and $K=1$. We identify the following important scales, see SM~\cite{SM}, section C for a detailed discussion: $\lambda_\alpha = \sqrt{K/\alpha}$ is the length scale over which gradients in $\boldsymbol{S}$ balance with deviations from $\boldsymbol{S}^2=1$, $\lambda_{\bar{\sigma}} = K/\bar{\sigma}$ is the length scale at which elastic and non-reciprocal forces balance, $\tau=\gamma/\alpha$ is the relaxation time of the system. Finally, $\sqrt{K\alpha}$ sets the non-reciprocity scale ($\bar{\sigma}$). Note that $\lambda_{\alpha}/\lambda_{\bar{\sigma}}   = \bar \sigma /\sqrt{K\alpha}$. We fix the system size to $L=2\pi$.
\newline

The essence of non-reciprocity on localized excitations is well captured by smooth, $1d$ perturbations. 
The smoothness assumption implies that gradients $|\nabla\boldsymbol{S}|$ are small. We can thus work under the approximation $S \approx 1$. The Landau term $\propto \alpha (1-S^2)$ becomes negligible and Eq.~(\ref{eq:main_eq}) can be expressed solely in terms of the orientation field $\theta$ : 
\begin{equation}
\gamma \,\dot \theta = K\Delta \theta + \bar{\sigma} (\nabla \times \hat{\textbf{S}})_z
= K\Delta \theta + \bar{\sigma} (\hat{\textbf{S}} \cdot \nabla)\theta \ , \\
\label{eq:main_eq_1dof}
\end{equation}
where $\hat{\textbf{S}} =(\cos \theta, \sin \theta)$. 
The $z$ index denotes the third component of the rotational (the only non-zero one). As we want to describe unidirectional propagation, we restrict Eq.~(\ref{eq:main_eq_1dof}) to the $x$-axis:
\begin{equation}
\gamma \,\dot \theta = \ K\theta_{xx} + \bar{\sigma} \cos (\theta) \,\theta_{x}  \   , 
\label{eq:main_eq_1dof_1d}
\end{equation}
where we defined $\theta_{x} = \partial \theta / \partial x$ and  $\theta_{xx} = \partial^2 \theta / \partial x^2$.
\ylann{The $\cos \theta$ term explicitly breaks the invariance under rotation. In our model, it follows from coupling local interactions to the local orientation. In other models, it can arise from the coupling of the phase dynamics to the spatial dynamics \cite{casagrande2026topology}}. 

\begin{figure}[t]
    \centering
    \includegraphics[width=\linewidth]{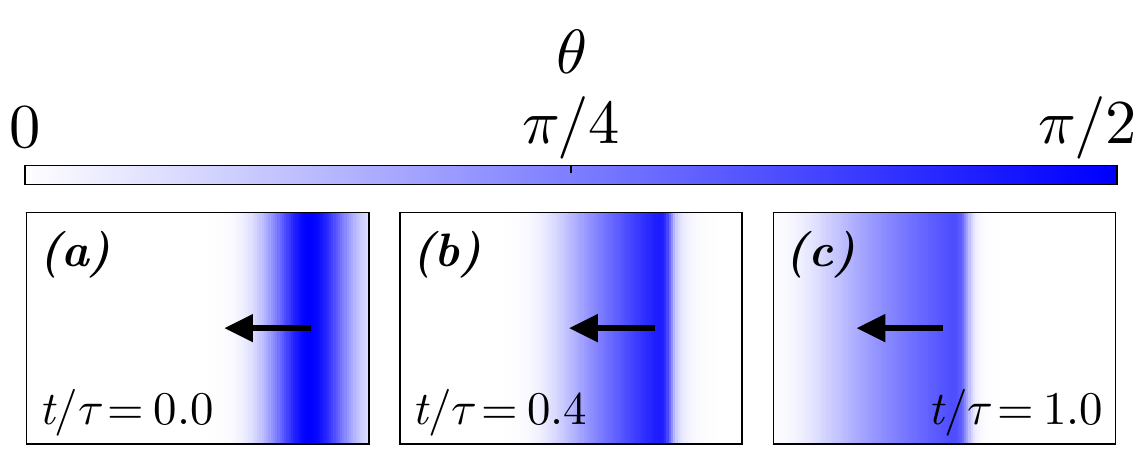}
    \caption{ Time evolution of an initial $1d$ gaussian excitation. We show a partial rectangular window of the system. 
    The orientation field $\theta$ is plotted in panels \textbf{(a-c)} for $t/\tau = 0, 0.4, 1$ respectively. The other parameters are $\alpha =100,\, \bar{\sigma}/\sqrt{K\alpha} = 20,\, \theta_0=0$. The initial profile travels to the left (black arrow) and develops a front/back asymmetry (smoother at the front, sharper at the back).}
    \label{fig:snapshot_propagation_1d}
\end{figure}

We study an idealized scenario and investigate the propagation of an initially localized perturbation $\delta$ along one direction:
\begin{equation}
    \theta(x, y, t) =\theta(x, t) = \theta_0 + \delta(x, t)\, . \\
  \label{eq:perturbation1}
\end{equation}
{Substituting this into Eq.~(\ref{eq:main_eq_1dof_1d}) leaves us with an equation for the evolution of the perturbation}
\begin{equation}
\gamma \,\dot \delta = \ K\delta_{xx} + \bar{\sigma} \cos (\theta_0+\delta) \,\delta_{x}  \   ,  
\label{eq:perturbation1_dynamics}
\end{equation}
{that highlights the role played by the different terms. The first diffusion-like term dissipates the perturbation while the second advection-like term causes the perturbation to move. We confirm this behavior by simulating Eq.~(\ref{eq:main_eq}) with $\theta_0=0$ and an initial Gaussian perturbation of the form:}
\begin{equation}
    \delta(x, t=0) = \delta_0 \exp\left(-\frac{(x - x_0)^2}{2 \, w_0^2}\right)  \ ,
  \label{eq:perturbation2}
\end{equation}
where $\delta_0$ is the height of the initial profile, $x_0$ its initial position, and $w_0\ll L$ its initial width. Figure.~\ref{fig:snapshot_propagation_1d} shows the time evolution of the $\theta$ field for a perturbation with $\delta_0 = \pi/2$ and $w_0/L = 0.05$. The perturbation spreads, develops a front/back asymmetry and propagates. We articulate the rest of this section around those three dynamical features. 

\begin{figure*}
    \centering
    \includegraphics[width=\linewidth]{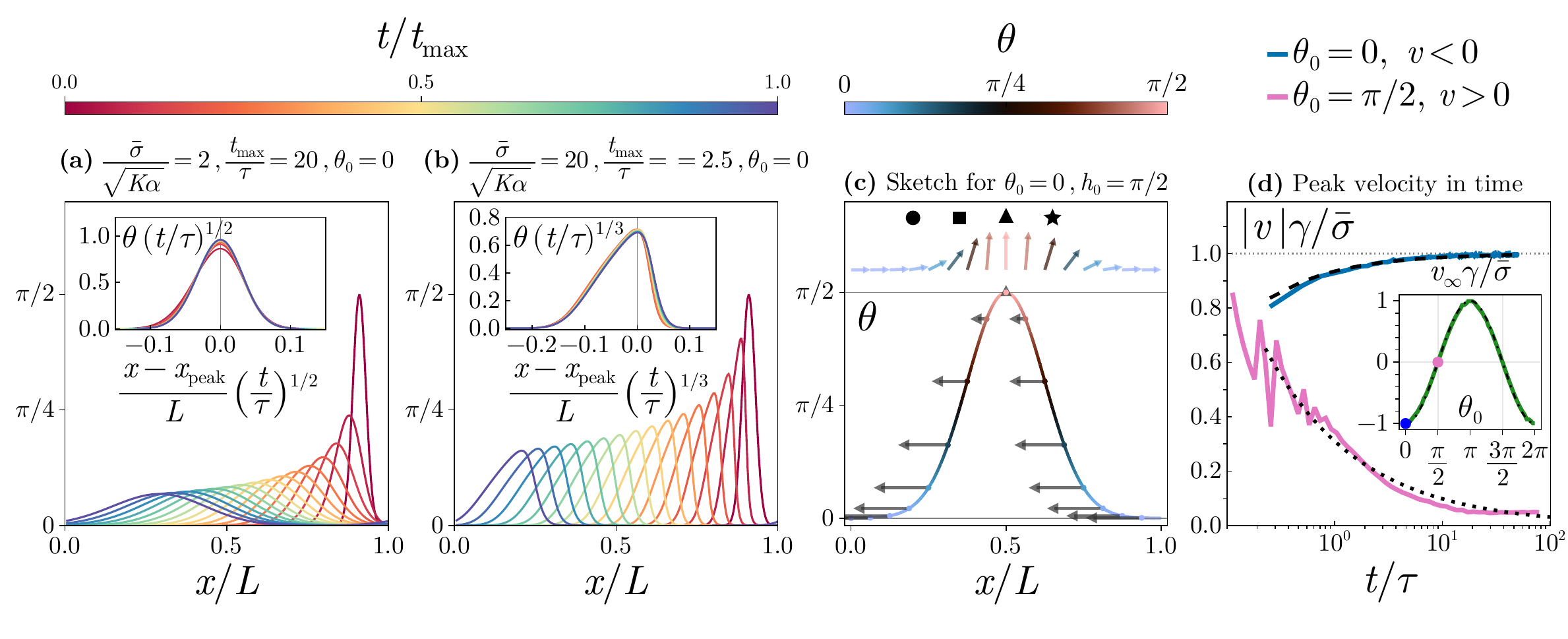}
    \caption{
    \textbf{(a)} Propagation of a perturbation $\theta=\delta(x)$ across the system over time. Parameters: small non-reciprocity $ \bar{\sigma}/\sqrt{K\alpha} = 2, \alpha = 100, t_{\text{max}}/\tau = 20, \theta_0=0$. \underline{Inset}: rescaled data $\theta\sqrt{t/\tau}$ as a function of $(x-x_{\text{peak}})/(L\sqrt{t/\tau})$.  
    \textbf{(b)} Same but for larger non-reciprocity $\bar{\sigma}/\sqrt{K\alpha} = 20$, and shorter simulation time $ t_{\text{max}}/\tau = 2.5$. \underline{Inset}: rescaled data $\theta \,(t/\tau)^{1/3}$ as a function of $(x-x_{\text{peak}})/(L(t/\tau)^{1/3}) $.    
          \textbf{(c)} A gaussian profile of height $\delta_0 = \pi/2$ and background orientation $\theta_0 = 0$. 
  Colors indicate the orientation $\theta$, see color map. The horizontal gray arrows represent $-\cos\theta$, which physically corresponds to the local convective velocity, see main text for details. The symbols at the top of the boxes are used in the main text to guide the reader. 
    \textbf{(d)} Peak velocity for $\theta_0=0$ (blue, $v<0$)  and $\theta_0=\pi/2$ (pink, $v>0$). 
          The dashed and dotted lines correspond to Eq.~(\ref{eq:sigmacos}). Dashed line: $\beta = 1/3, c= 0.72$.  Dotted line: $\beta = 1/2, c= 0.62$.  
     \underline{Inset}: steady peak velocity $v_\infty $ against $\theta_0$, for $\delta_0 = \pi/20$.  The back dash line perfectly matching the data is $-\cos \theta_0$.  }
    \label{fig:propagation}
\end{figure*}

\subsection{Spreading}

Equation~(\ref{eq:perturbation1_dynamics}) is non-linear due to the presence of the $\cos (\theta_0 + \delta)$ term. To gain insight on the spreading and propagation phenomena, we linearize Eq.~(\ref{eq:perturbation1_dynamics}) around $\theta_0$, considering that the excitation amplitude is small. The linearized equation can be easily solved in Fourier space, leading to the solution $\hat\delta(q, t)$ for wave number $q$ given by
\begin{equation}
    \hat\delta(q, t) \sim \exp\left(  -\frac{Kq^2}{\gamma} t + i\frac{\bar{\sigma}\cos \theta_0}{\gamma} q\,t \right) \ .
    \label{eq:fourier}
\end{equation}
These two terms highlight the damping and advection processes. 
The Laplacian damps each Fourier mode with a typical timescale $\gamma/Kq^2$, validating a \textit{posteriori} the small amplitude assumption. 
The second, imaginary term represents a translation, at a speed $v_0 = \bar{\sigma}\cos \theta_0/\gamma$. We shall revisit this when discussing propagation of excitations in the next section. To better rationalize the spreading process, we now go beyond a simple linear analysis.
The damping implies $\delta(t) \to 0 $ and justifies the expansion $\cos \theta \approx \cos\theta_0 \,(1-\delta^2/2)  - \sin \theta_0 \, \delta$, which allows to write Eq.~(\ref{eq:perturbation1_dynamics}) as a generalized Burgers equation~\cite{BEC20071}:
\begin{equation}
\dot \delta - v_0\,\delta_{x} \!= \!\frac{K}{\gamma }\delta_{xx} - \frac{\bar{\sigma} \sin \theta_0}{\gamma } \,\delta \,\delta_{x} - \frac{\bar{\sigma} \cos \theta_0}{2\gamma } \,\delta^2 \,\delta_{x}\   .
\label{eq:burgers}
\end{equation}
On the left hand side, the term $\propto \delta_{x}$ describes a rigid advection at speed $v_0$, a feature already captured by the linear Fourier analysis and that can be removed by working in a comoving reference frame. Importantly, the rigid advection speed $v_0$ does not depend on the amplitude $\delta$ of the perturbation and is therefore constant over time. 
On the right hand side, the first term is a diffusion term (or viscous term, in the terminology of the Burgers equation), while the two other terms are non-linear advection terms. 

Let us postulate that $\delta$ follows a universal scaling $\delta(x,t) \sim t^{-\beta} \, F(x/t^{\beta})$. The fact that the same power $\beta$ enters both inside and outside the function $F$ reflects the fact that both the width $w\sim t^{\beta}$ and the height $\delta_{\text{peak}}\sim t^{-\beta}$ of the perturbation are bound to evolve in the same fashion over time. Indeed, the mass under the curve $ M(t) = \int\delta(x,t) \, dx \sim w(t)\,\delta_{\text{peak}}(t)$ is conserved ($\dot M=0$, see SM~\cite{SM}, section D). 

This universal scaling implies $\delta_{x} \sim t^{-2\beta}$, $\delta_{xx} \sim t^{-3\beta}$, and $\dot \delta \sim t^{-\beta -1}$, see SM~\cite{SM}, section E. A simple dimensional analysis indicates three damping regimes, depending on which term dominates on the right hand side of Eq.~(\ref{eq:burgers}). If diffusion $\delta_{xx}$ dominates, i.e. {when $\lambda_{\bar{\sigma}} \gg \lambda_\alpha$}, one recovers $\beta = 1/2$, as expected. 
If non-reciprocity dominates over diffusion (when {$\lambda_{\bar{\sigma}} \ll \lambda_\alpha$}), two regimes are possible.  
{At short times, the $\delta^2\,\delta_{x}$ term dominates, leading to the scaling exponent $\beta = 1/3$. 
At long times, the $\delta\,\delta_{x}$ term dominates and we retain the $\beta = 1/2$ scaling. 
The crossover window between the two pure scaling regimes 
extends from $\tau \, \zeta^2$ to $\tau \, \zeta^3$, where $\zeta~\sim~\min \Big \{1/\tan \theta_0, \cos \theta_0  \cdot \bar{\sigma}/\sqrt{K\alpha} \Big \}$ 
can be controlled by changing the parameters of the model, in particular $\bar{\sigma}$ and $\theta_0$, see SM~\cite{SM}, section E for details.}

Motivated by these theoretical predictions, we plot in Fig.~\ref{fig:propagation}(a,b) the time evolution of the perturbation, with $\theta_0 = 0$ and two non-reciprocity values: small $\bar{\sigma}$ in panel (a) and large $\bar{\sigma}$ in panel (b); different time points are plotted in different colors, from red ($t=0$) to blue ($t = t_\text{max}$). We numerically identify the peak position $x_{\text{peak}}(t)$ of the perturbation at each time and recenter the profiles on it, in the spirit of a comoving frame. The insets show the same data but rescaled according to the $\beta$ values predicted above.
For small $\bar{\sigma}$, the crossover time is small and the elasticity dominates. The profiles collapse onto a single gaussian master curve when rescaling both axes according to $\beta = 1/2$, confirming the symmetric, diffusive nature of the spreading. For larger $\bar{\sigma}$ the crossover time is longer than the time scale explored in our simulations: we access the non-reciprocity dominated regime. In this regime one has to rescale the axes according to $\beta=1/3$ for the curves to collapse on a single, asymmetric master curve. 

\subsection{Asymmetry}  
As illustrated in Fig.~\ref{fig:snapshot_propagation_1d}, 
perturbations are indeed typically skewed: the front end gets stretched while the trailing end gets compressed.
This asymmetry arises because the advecting velocity depends on the profile itself. Indeed,  the advective velocity $v = \bar{\sigma} \, \cos (\theta_0+\delta(x,t))\,/\,\gamma$ is both space and time-dependent. Thus different parts of the perturbation are advected at different speeds.
We illustrate this in Fig.~\ref{fig:propagation}(c), where the modulus of the grey arrows correspond to the magnitude of $\cos \theta(x)$. In this sketch, $\theta_0=0$ and the height of the perturbation is $\delta_0=\pi/2$. The local advection speed is maximal at the base of the perturbation and minimal at the peak.
This creates a \emph{velocity shear}: the base is advected faster than the peak, hence the profile stretches on the left and gets compressed on the right, skewing the initially symmetric shape.  Elasticity maintains a coherent, steady shape, confirmed by the  curve collapse in the insets of Fig.~\ref{fig:propagation}(a,b). 

{In the long time limit, the system is dominated by the generalized elasticity of the field, and the perturbation is small. In this regime, our linear analysis is correct and the asymmetry is lost. However, in the non-reciprocity dominated regime, the} asymmetry is captured by {Eq.~(\ref{eq:burgers}), similar to} a viscous Burgers equation, a paradigmatic minimal framework to describe shock-waves \cite{burgers_review}. {In Eq.~(\ref{eq:burgers}) there are three advection terms which combine to describe how the local  velocity depends on the local perturbation $\delta$, and on the background orientation $\theta_0$. The two non-linear advection terms are antisymmetric around the perturbation, hence they drive the asymmetry by acting differently on its leading/trailing edge.} 
We provide in SM~\cite{SM}, section F various plots illustrating the motion of the perturbation for various $0\le \theta_0 < 2\pi$, including configurations where the advection velocity changes sign through the perturbation. This happens when $\cos(\theta_0)$ and $\cos(\theta_0+\delta)$ have different sign, for example if $\theta_0 < \pi/2$ and $\theta_0+\delta>\pi/2$. This results in a perturbation in which the base and the tip of the excitation travel in opposite directions for a transient period.  
\newline 

\subsection{Propagation }
Before diving into the quantitative characterization of the propagation of an excitation, let us provide a first simple description based on the vision cone microscopic point of view.  To guide the reader through the following paragraph, we have drawn in Fig.~\ref{fig:propagation}(c), 
in addition to the profile $\theta(x)$, the corresponding arrows and four symbols (circle $\bullet$, square $\blacksquare$, triangle $\blacktriangle$ and star $\star$). 
Let us first consider the spin $\blacksquare$. It points slightly to the right, so its vision cone makes it more sensitive to the spin $\blacktriangle$ than to the spin $\bullet$. It will preferentially align with $\blacktriangle$ and thus rotate counterclockwise. On the other hand, the spin $\star$ looks away from the perturbation and will therefore rotate clockwise to align with the horizontal background. 
Altogether, the peak of the profile displaces to the left, in the direction opposite to the background medium orientation.\\ 

A visual inspection of the data, eg. Fig.~\ref{fig:snapshot_propagation_1d} or Fig.~\ref{fig:propagation}(b), reveals that perturbations can be long-lived and travel over distances of the order of $10^3\,\lambda_\alpha$ before ``vanishing" (since the amplitude of the perturbation decays following a power-law, there is no natural scale to define a proper vanishing time).
A similar phenomenon was observed in an agent-based model with a sharp vision cone and Glauber dynamics \cite{loos2023long}. This confirms that the propagation of excitations is robust and general, thus calling for a more thorough investigation. 
\newline

{We have shown that after you account for spreading and dissipation, the perturbation is advected with a constant shape, see Fig.~\ref{fig:propagation}(a,b). The velocity of the whole perturbation is thus fully described by the velocity of the peak, $v_{\text{peak}}$. Gathering all the advection terms in Eq.~(\ref{eq:burgers}) gives a local advection velocity of}
\begin{equation}
    v = \frac{\bar{\sigma}}{\gamma}\left(\cos\theta_0\left(1-\frac{\delta^2}{2}\right) - \delta\sin\theta_0\right).
        \label{eq:v_cos_sin_taylor}
\end{equation}
Since $v_{\text{peak}}$ is the collective velocity of the whole perturbation, we expect it to depend on the average amplitude of the perturbation, which we crudely approximate to the half-amplitude $\theta_{1/2} = \theta_0  + \delta_{\text{peak}}/2 $. 
{We finally use the scaling $\delta_{\text{peak}}=c\,(t/\tau)^{-\beta}$ derived from the generalized Burgers equation. The coefficient $c$, extracted from the data, can be read from the maximum value of the curves in the insets of Fig.~\ref{fig:propagation}(a,b)}.  We obtain
\begin{equation}
    \frac{\gamma \,v_{\text{peak}}}{\bar \sigma} = \cos \theta_0 \left[1 - \frac{c^2}{8}\left(\frac{t }{\tau}\right)^{2\beta}\right] - \sin \theta_0 \ \frac{c}{2}\left(\frac{t }{\tau}\right)^\beta\,\ .
    \label{eq:sigmacos}
\end{equation}
Equation~(\ref{eq:sigmacos}) indicates two possible regimes, depending on the orientation $\theta_0$ of the background medium. 
If $\cos \theta_0 \neq 0$, the background medium supports constant advection with velocity $v_0 = \bar{\sigma} \, \cos (\theta_0)\,/\,\gamma$, as predicted by the linear Fourier analysis. As the system evolves, $\delta \to 0$ and $v_{\text{peak}}\to v_0$, see Fig.~\ref{fig:propagation}(d). For $\theta_0=0$ (blue curve), the background medium points to the right, which explains why the excitation travels to the left ($v<0$), in direction opposite to $\boldsymbol{\hat S}_0=(\cos\theta_0,\sin\theta_0)$. 
The behavior of $v_{\text{peak}}$ is well captured by Eq.~(\ref{eq:sigmacos}) (see dash lines Fig.~\ref{fig:propagation}(d)), confirming that the perturbation velocity indeed depends on its average amplitude rather than on its peak amplitude. 
We report in the inset of Fig.~\ref{fig:propagation}(d) the $\cos \theta_0$ modulation of the steady velocity of its peak. For all $\theta_0\neq\pm\,\pi/2$, the orientation of the background medium has a non-zero component along $x$ and can therefore sustain a propagation of the excitation. 

On the other hand, no long-lived propagation can be sustained in the direction perpendicular to $\boldsymbol{\hat S}_0$ (when $\theta_0 = \pm\, \pi/2$).
This can be understood either from the kernel perspective [$g(\varphi = +\pi/2) = g(\varphi = -\pi/2) = 1$ implying no net information flux] or from Eq.~(\ref{eq:sigmacos}) [where the comoving frame velocity $v_0\sim \cos \theta_0$ vanishes for $\theta_0 = \pm\, \pi/2$]. 
{In this scenario, the advective force only acts where $\theta\neq\theta_0$, ie. within the perturbation. The advection velocity is maximal at the peak of the perturbation, which causes the peak to internally displace, generating asymmetry. This creates higher local gradients $\nabla \delta$, that the elastic Laplacian dissipates isotropically; causing the whole perturbation to move. Thus propagation is powered from the inner non-reciprocal active force, and kept in a coherent shape by the reciprocal elastic force. These dynamics are well described by the $\sin\theta_0$ term in Eq.~(\ref{eq:sigmacos}). The direction and speed of propagation now depends on the sign and magnitude of $\delta$, respectively. In our simulations, $\delta>0$,  thus, the perturbation propagates to the \textit{right} and the speed quickly decays as the perturbation dissipates, see the pink decaying curve in Fig.~\ref{fig:propagation}(d), consistent with the long term velocity $\sim \cos\theta_0=0$ (pink dot in the inset of Fig.~\ref{fig:propagation}(d)).
While the $\theta_0=\pm\,\pi/2$ limit case is singular in $1d$, it will nevertheless prove useful in describing the motion of a perturbation in $2d$.

\begin{figure}[b!]
    \centering
    \includegraphics[width=\linewidth]{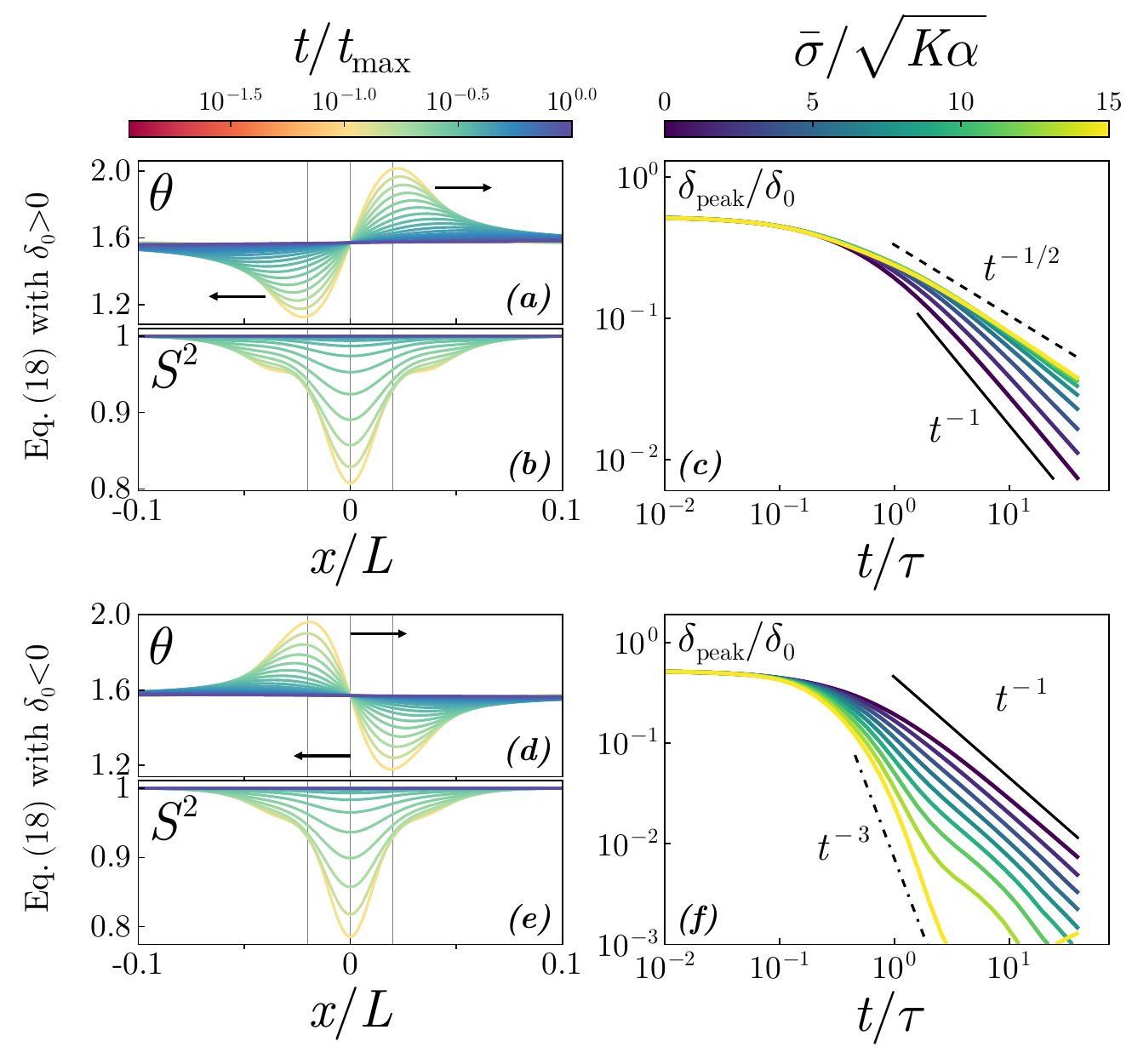}
    \caption{
    \textbf{(a)} Profile $\theta(x) = \pi /2 + \delta$, where $\delta$ is given by Eq.~(\ref{eq:sinx_1d}) with $\delta_0>0$. Different colors are different times on a logarithmic scale, increasing from yellow to blue (steady state). The same color scale is used in panels (a,b,d,e). The vertical grey lines are plotted to guide the eye and have the same value in all panels (a,b,d,e). The arrows represent the direction of motion of the two parts of the perturbation. 
    \textbf{(b)} Same plot for the squared magnitude $S^2(x)$. It dips at $x=0$ where the derivative $\partial\theta/\partial x$ is the largest. 
    \textbf{(c)} The relative amplitude $\delta_\text{peak}/|\delta_0|$ of a perturbation with $\delta_0>0$ as a function of time for different values of non-reciprocity. The initial value $\delta_\text{peak}/\delta_0 = 0.523$. The dashed and solid lines indicate $\sim t^{-1/2}$ and $\sim t^{-1}$, respectively.
    \textbf{(d-f)} Same quantities but for  $\delta_0<0$, swapping the locations of the two peaks.
    In panel (f), the solid and dotted lines indicate $\sim t^{-1}$ and $\sim t^{-3}$, respectively.}
    \label{fig:propagation_dan_1d}
\end{figure}
Finally, advection driven by the perturbation itself rather than the background medium relies on the perturbation breaking left-right symmetry. If $\delta$ is odd ($\delta(-x) = -\delta(x)$), the positive and negative parts of the perturbation propagate in opposite directions and cancel each other out when $\theta_0=\pm\,\pi/2$.
We report in Fig.~\ref{fig:propagation_dan_1d} the dynamics of an odd $1d$ perturbation of initial shape 
\begin{equation}
    \delta(x, t=0) = \delta_0 \sin(x/w_0) \, \exp\left( -\frac{x^2}{2w_0^2}\right) \ ,
    \label{eq:sinx_1d}
\end{equation}
evolving in a $\theta_0 = \pi/2$ background. 
We first focus on the $\delta_0>0$ case.
We plot the time evolution of the orientation $\theta(x)$ in Fig.~\ref{fig:propagation_dan_1d}(a) and of the magnitude $S^2(x)$ in Fig.~\ref{fig:propagation_dan_1d}(b). Different colors represent different times, from intermediate  (yellow) to steady (blue) values. The profile remains odd around the fixed point between the two peaks located at $x=0$. The total mass of the perturbation is again conserved and remains zero throughout. In addition, there is no net motion of the perturbation, with the third moment of the distribution remaining zero as well. 

Due to the additive nature of $\delta$, the positive and negative parts of the perturbation cancel each other out when they meet, leading to a faster decay of the perturbation. 
To provide analytical insight, we first consider the case where $\bar\sigma=0$ and the dynamics are governed by a $1d$ diffusion equation. The solutions for the diffusion equation are given by the convolution of the initial conditions with the Green's function, $G(x,t)$; for the diffusion equation $G(x,t)$ is a Gaussian. For an odd initial condition such as that studied in this section, the zeroth moment of the profile (the total mass) is strictly zero and the long time dynamics are dominated by the dipole moment of the distribution. This implies that the peak absolute magnitude of the profile dissipates according to $\delta_\text{peak}\sim t^{-1}$, see SM~\cite{SM}, section G for details. This describes the dynamics of our odd state Eq.~(\ref{eq:sinx_1d}) in the absence of non-reciprocity, see dark blue curves and solid black line in Fig.~\ref{fig:propagation_dan_1d}(c).

When non-reciprocity is introduced, $\bar \sigma>0$, the positive part of the perturbation ($\theta > \pi/2, \delta>0$) propagates to the right fueled by internal dynamics. Conversely, the negative part propagates to the left leading to diverging motion of the two peaks indicated by the arrows in Fig.~\ref{fig:propagation_dan_1d}(a). The diverging motion of the peaks leads to reduced interaction between the positive and negative parts of the perturbation and the decay is slowed down. For sufficiently strong non-reciprocity, the peaks become independent, recovering the behavior of an isolated peak, i.e. $\delta_\text{peak}\sim t^{-1/2}$, see yellow curves in Fig.~\ref{fig:propagation_dan_1d}(c). 

In the $\delta_0<0$ case, the position of the peaks are switched and the divergent motion becomes convergent motion, see Fig.~\ref{fig:propagation_dan_1d}(d). This increases the interaction between the positive and negative parts of the perturbation and causes the dissipation to disperse at a superdiffusive rate as non-reciprocity is increased (up to $\delta_\text{peak}\sim t^{-3}$), see Fig.~\ref{fig:propagation_dan_1d}(f). Thus non-reciprocity either increases or decreases the dissipation rate of a stationary, odd perturbation depending on the sign of the perturbation. 

\subsection{Propagation of a $2d$ perturbation}

\begin{figure}[b!]
    \centering
    \includegraphics[width=\linewidth]{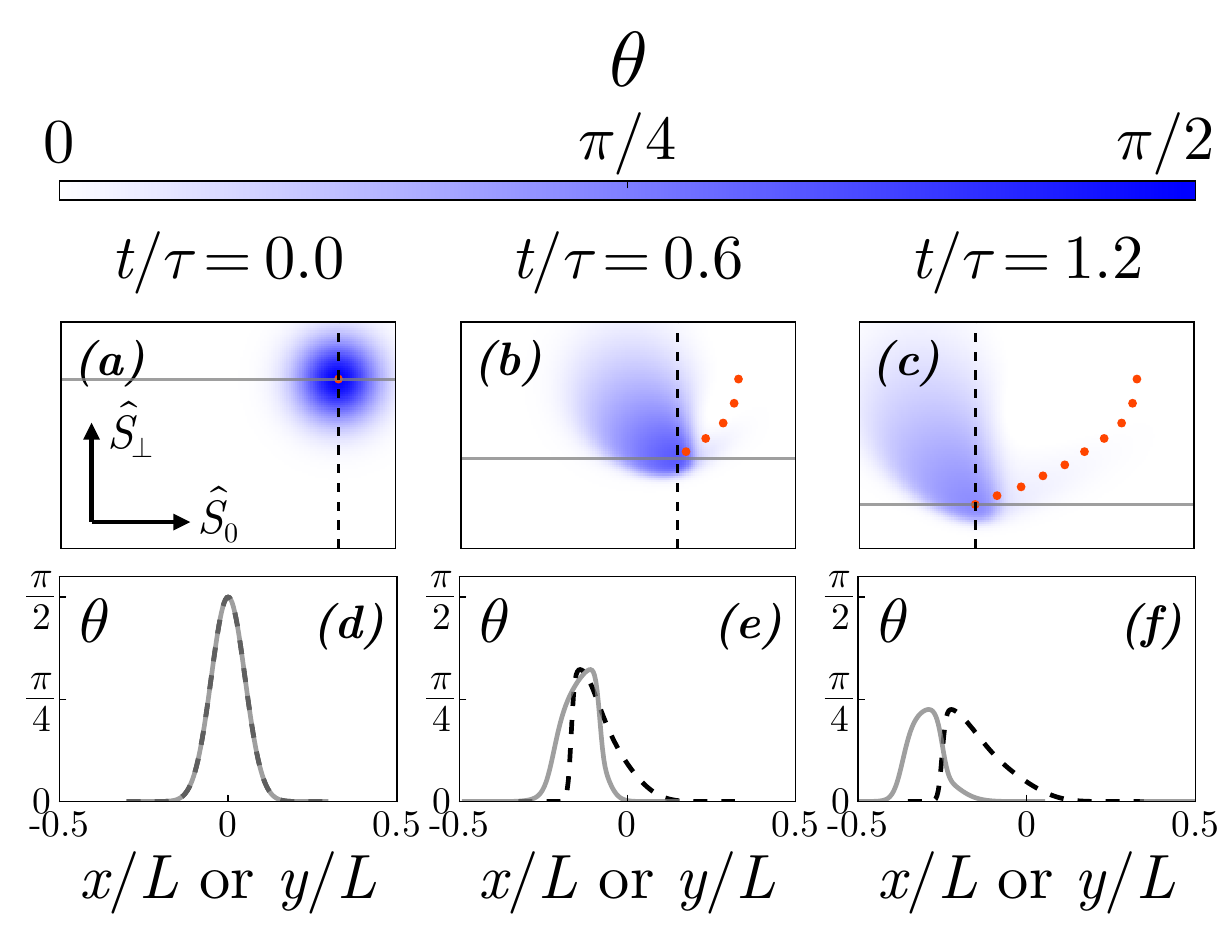}
    \caption{ Time evolution of an initial isotropic $2d$ gaussian perturbation, with $\theta_0 = 0, \delta_0 =\pi/2$. We show a partial rectangular window of the system. 
    The orientation field $\theta$ is plotted in panels \textbf{(a-c)} for $t/\tau = 0, 0.6, 1.2$ respectively. The dots track the past positions of the peak. Other parameters: $\alpha =100, \bar{\sigma}/\sqrt{K\alpha} = 20$. \ylann{$\mathbf{\hat S}_0 = (\cos \theta_0, \sin \theta_0)$ and $\mathbf{\hat S}_\perp = (-\sin \theta_0, \cos \theta_0)$. }
    \textbf{(d-f)}  cross-sections of the orientational field $\theta$ along a vertical (dash, against $y/L$) and horizontal (solid, against $x/L$) lines passing through the peak of the perturbation in panels (a-c). 
  }
    \label{fig:snapshot_propagation_2d}
\end{figure}

\begin{figure*}
    \centering
    \includegraphics[width=\linewidth]{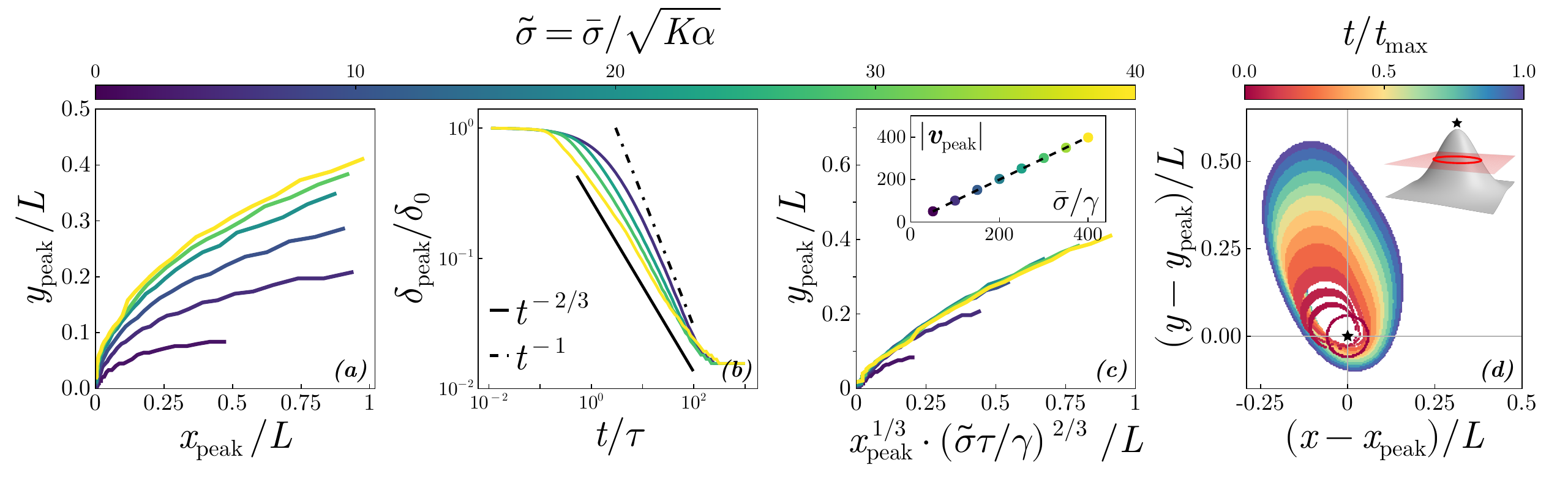}
    \caption{
    \ylann{
    \textbf{(a)} 
    Variation of the trajectory $ y( x)$ (here $\theta_0 = \pi$) with different $\bar{\sigma}$, corresponding to different colors. We cut the curves before $x = L$, so the time span of each curve is different (it decreases with $\bar \sigma$).    \textbf{(b)} Decay of the normalized peak amplitude over time. The black dash line is $\sim t^{-1}$, the black solid line is $\sim t^{-2/3}$. 
    \textbf{(c)} Same data as in panel (a), but on rescaled axes, see main text. 
    \underline{Inset}: Scalar velocity of the peak. The dash line is of slope 1.
    \textbf{(d)} Flattening and asymmetry of the profile over time, for $\bar{\sigma}/\sqrt{K\alpha} =40$. For different times [in different colors, from red ($t=0$) to blue ($t_\text{max} = 5\tau$)], we plot the mid-height manifolds (red in the sketch), centered on the peak (black star). 
    For all 4 panels: the initial height, we use  $\alpha = 100$ and take an initial perturbation as in Eq.~(\ref{eq:2d_perturbation}), with $\delta_0 = \pi/2 $. 
    }
    }
    \label{fig:propagation2d}
\end{figure*}
We now expand the discussion to a $2d$ perturbation, and first take the example of an isotropic gaussian initial perturbation
\begin{equation}
    \delta( x,y,t=0) = \delta_0 \exp\left(-\frac{x^2 + y^2}{2\,w_0^2}\right) 
    \label{eq:2d_perturbation}
\end{equation}
where $\delta_0$ is the height of the initial profile, and $w_0\ll L$ its initial width (the same along $x$ and $y$). We use $\delta_0 = \pi/2$ and $w_0/L = 0.05$, as for the $1d$ perturbation. 
A change in $\theta_0$ can be mapped to a global rotation of the system now that the perturbation at $t=0$ is a rotationally invariant  gaussian.
We first take $\theta_0 = 0$ and show snapshots of the perturbation over time in Fig.~\ref{fig:snapshot_propagation_2d}(a-c), with the trajectory of the peak plotted with red dots. 
The perturbation takes a curved path. 
In the bottom row, we plot the cross-sections of the orientational field $\theta$ along a vertical (dash) and horizontal (solid) lines passing through the peak of the perturbation. 
Even though the profiles are different from the $1d$ excitation case, the three main features remain: its propagation,  flattening and skewing. 

\ylann{

We now choose $\theta_0 = \pi$ for convenience, such that the perturbation travels from left to right. We monitor the displacement of the peak over time and report the excitation trajectory in Fig.~\ref{fig:propagation2d}(a) for different non-reciprocity values. We have checked that varying $\alpha$ does not change the results, consistent with the fact that $S\approx 1$ everywhere.

To rationalize the curved path of the excitation and the decay of its amplitude $\delta_{\text{peak}}$, see Fig.~\ref{fig:propagation2d}(b),  we extend the analysis of the Burgers equation to $2d$. Under the assumption of a small perturbation $\delta$, and with $\mathbf{\hat S}_0 = (\cos \theta_0, \sin \theta_0)$ and $\mathbf{\hat S}_\perp = (-\sin \theta_0, \cos \theta_0)$, we arrive to 
\begin{equation}
\gamma \dot \delta = K\nabla^2\delta  
+ \bar \sigma \Big(\mathbf{\hat S}_0 \cdot \nabla\delta\Big)
+  \bar \sigma  \Big(\mathbf{\hat S}_\perp \cdot \nabla\delta\Big) \delta
- \frac{\bar \sigma^2}{2} \Big(\mathbf{\hat S}_0 \cdot \nabla\delta\Big) \delta^2
    \label{eq:burgers_2d}
\end{equation}
We again postulate that the perturbation follows a universal scaling, which in this case reads $\delta(x,y,t) \sim t^{-\beta} F(x/t^{\beta/2} , y/t^{\beta/2})$, and conduct the same dimensional analysis as in the $1d$ case. When the diffusive terms dominate, one obtains $\beta=1$, as expected for $2d$ diffusion, see the dash-dotted lines in Fig.~\ref{fig:propagation2d}(b). When the active terms dominates, two regimes are possible. If the dominant term is $\delta^2$, one obtains $\beta = 2/5$. We do not observe this decay, probably because the decay exponents in $2d$ are greater than in $1d$, such that $\delta^2$ becomes rapidly negligible before $\delta$. In this regime, one has $\beta = 2/3$, which we observe over two decades, see Fig.~\ref{fig:propagation2d}(b). 

We now address the curved shape of the trajectory of the peak $(x_{\text{peak}}, y_{\text{peak}})$ of the perturbation . The first two advection terms in Eq.~(\ref{eq:burgers_2d}) advect $\delta$ in the directions $-\mathbf{\hat S}_0$ and $-\mathbf{\hat S}_\perp$ respectively, as illustrated in  Fig.~\ref{fig:snapshot_propagation_2d}. For the case $\theta_0=\pi$ in Fig.~\ref{fig:propagation2d}(b,c), $-\mathbf{\hat S}_0$ and $-\mathbf{\hat S}_\perp$ are along the $x$ and $y$ axes, respectively. The advection $\mathbf{\hat S}_0 \cdot \nabla\delta$ in the $x$ direction is constant over time as it does not depend on the amplitude $\delta$. This yields 
\begin{equation}
    \dot x_{\text{peak}} = \frac{\bar{\sigma}}{\gamma}\longrightarrow x_{\text{peak}} = \frac{\bar{\sigma} \,t}{\gamma} \ .
    \label{eq:xpeakt}
\end{equation}
On the contrary, the dynamics in the $y$ direction depend on the amplitude $\delta$, which we now know scales with $t^{-\beta} = t^{-2/3}$ for a large window of intermediate times. The equation for  $ \dot y_{\text{peak}}  $ thus reads
\begin{equation}
    \dot y_{\text{peak}}  
    \sim
    \frac{\bar{\sigma}\,\delta_\text{peak}}{\gamma}
    \sim 
    \frac{\bar{\sigma}}{\gamma}\,\left(\frac{t}{\tau}\right)^{-2/3} \ ,
\end{equation}

and thus 

\begin{equation}
   y_{\text{peak}}\sim \frac{\bar{\sigma}\, t^{1/3}\, \tau^{2/3}}{\gamma}= \left(\frac{\bar{\sigma} \,\tau }{\gamma }\right)^{2/3}\, x_{\text{peak}}^{1/3} \ .
        \label{eq:ypeakt}
\end{equation}

} 

\begin{figure*}
    \centering
    \includegraphics[width=\linewidth]{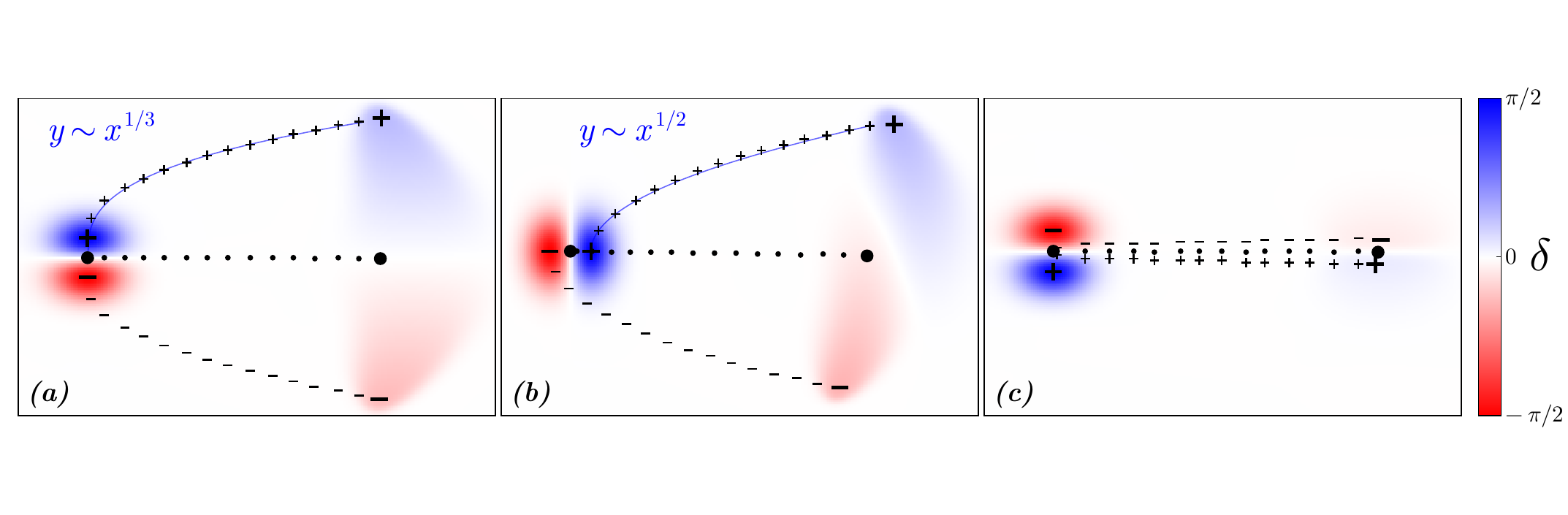}
    \vspace{-1.3cm}
    \caption{ Different initial conditions lead to qualitatively different trajectories. We illustrate three cases for the ~vector $\bm{k}$ in Eq.~(\ref{eq:2d_AS_perturbation}) : 
    \textbf{(a)} $\bm{k} = [1/w_0,0]$ 
    \textbf{(b)}  $\bm{k} = [0,1/w_0]$
    \textbf{(c)} $\bm{k} = [-1/w_0,0]$. 
    In each panel, we superimpose the initial state (on the left, well defined lobes with intense colors) and the state at $t_\text{max}/\tau = 1.45$ (on the right, with faded colors due to spreading). The $\textbf{+}$ (resp. $\textbf{-}$) symbols correspond to the location of the global maximum (resp. minimum) of $\delta(x,y)$ at different times. The circles ($\bullet$) correspond to the trajectory of the barycenter of the excitation. 
    For all panels, $\alpha = 1000, \bar \sigma = 300, \theta_0=\pi, w_0/L =  0.05, |\delta_0/0.523|=\pi/2$, and the value of $\delta = \theta - \theta_0$, from $-\pi/2$ to $\pi/2$ is represented on the color scale on the right.  
    }
    \label{fig:propagation_dan_2d}
\end{figure*}
The last equality is obtained by isolating $t$ in Eq.~(\ref{eq:xpeakt}). We plot in Fig.~\ref{fig:propagation2d}(c) the same data as in Fig.~\ref{fig:propagation2d}(b) but with the $x$ axis rescaled as suggested by Eq.~(\ref{eq:ypeakt}). The curves for different non-reciprocity levels collapse, confirming that our $1d$ results can indeed be extended to a $2d$ description. In the same spirit, we report in the inset of Fig.~\ref{fig:propagation2d}(c) that the norm of the velocity of the peak $|{\textbf{v}}_{\text{peak}}|$ is equal to $ \bar \sigma / \gamma$, as in the $1d$ case. 

Finally, we look at the deformation of the perturbation over time. In the spirit of the width at mid-height for 1$d$ profiles, we extract the mid-height manifold and present the results in Fig.~\ref{fig:propagation2d}(d) for different times (different colors). More precisely, we center the data such that the peak of the perturbation is at the origin $(0,0)$ at all times and we plot all the spins $(x,y)$ such that $|\theta(x,y) - \frac{1}{2}\,\underset{x,y}{\max} \ \theta(x,y,t)| \le  0.04$ (within a small, constant and arbitrary margin 0.04).

{
The initially isotropic gaussian evolves into an anisotropic, asymmetric shape, mirroring the loss of symmetry observed in the $1d$ case. Again we can shed some light on this process by noticing that the change of shape parallel or perpendicular to the background orientation can be interpreted by considering a $1d$ perturbation with $\theta_0 = 0$ or $\pi/2$, respectively. From our $1d$ analysis, we expect spreading perpendicular to the background orientation to be faster than spreading parallel, which leads to the perturbation becoming elongated in the $y$ direction, see Figs.~\ref{fig:snapshot_propagation_2d}(f)~and~\ref{fig:propagation2d}(d). Similarly, from our $1d$ analysis we expect the trailing edge of a perturbation moving parallel to the background orientation to be steeper, whereas the opposite is true for a perturbation moving perpendicular to the background, see SM~\cite{SM}, section F (Fig.~1 therein). A similar effect is reflected in the asymmetry of a $2d$ perturbation, see Fig.~\ref{fig:snapshot_propagation_2d}(f).}

The trajectory of a $2d$ perturbation is not constrained  to follow the $y\sim x^{1/3}$ predicted earlier. In fact, its fate is also controlled by the initial condition.
Extending our results on odd $1d$ excitations to the $2d$ case, we show we can remove the net motion of the perturbation in the $y$ direction. As in $1d$, we modulate our perturbation to introduce asymmetry along a wave vector $\bm{k}=[k_x, k_y]$.  Our new perturbation becomes
\begin{equation}
    \delta( x,y,t=0) = \delta_0 \sin(xk_x + yk_y)\exp\left(-\frac{x^2 + y^2}{2\,w_0^2}\right).
    \label{eq:2d_AS_perturbation}
\end{equation}

Equation (\ref{eq:2d_AS_perturbation}) describes perturbations that feature two prominent lobes of opposite sign. We present trajectories for the three extrema values of $\bm{k}$ in Fig.~\ref{fig:propagation_dan_2d} with a $\theta_0=\pi$ background. In each panel, we display the initial configuration on the left (two well defined lobes, red for negative $\delta$ and blue for positive $\delta$) together with its state after a given duration (two faded halos). The $\textbf{+}$ (resp. $\textbf{-}$) symbols correspond to the location of the global maximum (resp. minimum) of $\delta(x,y)$ for different times. The circles ($\bullet$) correspond to the barycenter of the two lobes, highlighting the path taken by the excitation. All perturbations have net motion to the right due to advection by the background medium ($\theta_0=\pi$). In all cases, the positive lobe tends to travel up, the negative lobe tends to travel down and the center of mass of the perturbation remains strictly on the $x$ axis, respecting the symmetry of the initial configuration.

For perturbations odd in the $+y$ axis ($\bm{k} = [0,1/w_0]$), the lobes move apart from each other in a symmetric fashion, see Fig.~\ref{fig:propagation_dan_2d}(a). This is similar to the $1d$ case shown in Fig.~\ref{fig:propagation_dan_1d}(a). This minimizes interactions between the two lobes and their individual trajectories are well described by Eq.~(\ref{eq:ypeakt}). For perturbations odd in the $+x$ axis $\bm{k} = [1/w_0,0]$, see Fig.~\ref{fig:propagation_dan_2d}(b), the positive and negative lobes move past each other. Interactions between the lobes cause them to locally dissipate faster. Since this effect is maximized where the lobes are closest, the resulting peaks migrate apart at an increased rate, which is fitted well by $y\sim x^{1/2}$. Finally, for perturbations odd in the $-y$ axis ($\bm{k} = [0,-1/w_0]$), see Fig.~\ref{fig:propagation_dan_2d}(c), the lobes have convergent dynamics similar to that shown in Fig.~\ref{fig:propagation_dan_1d}(d). This yields an excitation that travels along the $x$ axis, remains spatially confined in the $y$ axis, and decays faster. We thus show that one can render significant control over the propagation of a given perturbation via the design of its initial shape and background.
\newline

In this section, we have shown how the non-reciprocal active force advects and reshapes perturbations. 
We have first focused on $1d$ perturbations, and we have shown that a linear Fourier analysis of our dynamical equation captures the damping and the constant advection of the background medium. Going beyond the linearized regime, we have related our continuum model to a generalized Burgers equation. This minimal PDE appears in a variety of contexts, from $1d$ turbulence to shock-waves dynamics.  It captures the essence of the impact of non-reciprocity on perturbation dynamics and accounts for the dynamics of the first three moments of the profile. Excitations propagate in the direction opposite to that of the background medium (displacement of the mean), spread over time (increase of the standard deviation) and develop a front/rear asymmetry (non-zero skewness). 
We mention here that higher moments do not appear to play any role. One can reproduce the exact same results with an initial quartic profile $\propto\exp(-x^4)$: after a very short transient time, the profile is indeed indistinguishable from the initial gaussian case, see SM~\cite{SM}, section H. \\
We have then extended our results to $2d$ perturbations, showing that they abide by the same general principles. We have finally used this knowledge and symmetry considerations to design initial excitation shapes that realize specific trajectories.  \\ 

\section{3. Extended topologically protected excitations}

We now turn to a different kind of excitation that is not localized but spans the whole system: a modulation of the orientation field with a winding number equal to $k$ (ie. the orientation field $\theta$ wraps around $2\pi$ $k$ times). This is a configuration we refer to as a topologically protected spinwave. A simple linear spinwave is given by the following equation
\begin{equation}
    \theta(x,y) = \theta_0 + 2\pi k\,x/L \text{ and } S(x,y)=1 \ .
    \label{eq:spinwave}
\end{equation}
Such a configuration is apolar: {$P = |\int \exp(i\theta)d^2x|/L^2=0$}.
We represent a linear spinwave with winding number $k=1$ in Fig.~\ref{fig:snapshots_concentration}(a). In the limit $L \to \infty$, the elastic cost to sustain a small angular difference $\varepsilon = 2\pi/L$ between two neighbors is $\sim \varepsilon^2 \sim L^{-2}$ so the energy of this pattern is finite in the thermodynamic limit.

In this section, we study the impact non-reciprocity has on such a pattern.  Since it is topologically protected, it cannot smoothly vanish like the localized $1d$/$2d$ {gaussian} profiles of the last section did. As before, the curl term in Eq.~(\ref{eq:main_eq}) breaks global rotation invariance; however, for this specific profile combined with periodic boundary conditions, a global rotation is equivalent to a global translation along $x$, thus without loss of generality we set $\theta_0=0$ and consider the profile of Eq.~(\ref{eq:spinwave}).

\begin{figure}[b]
    \centering
    \includegraphics[width=\linewidth]{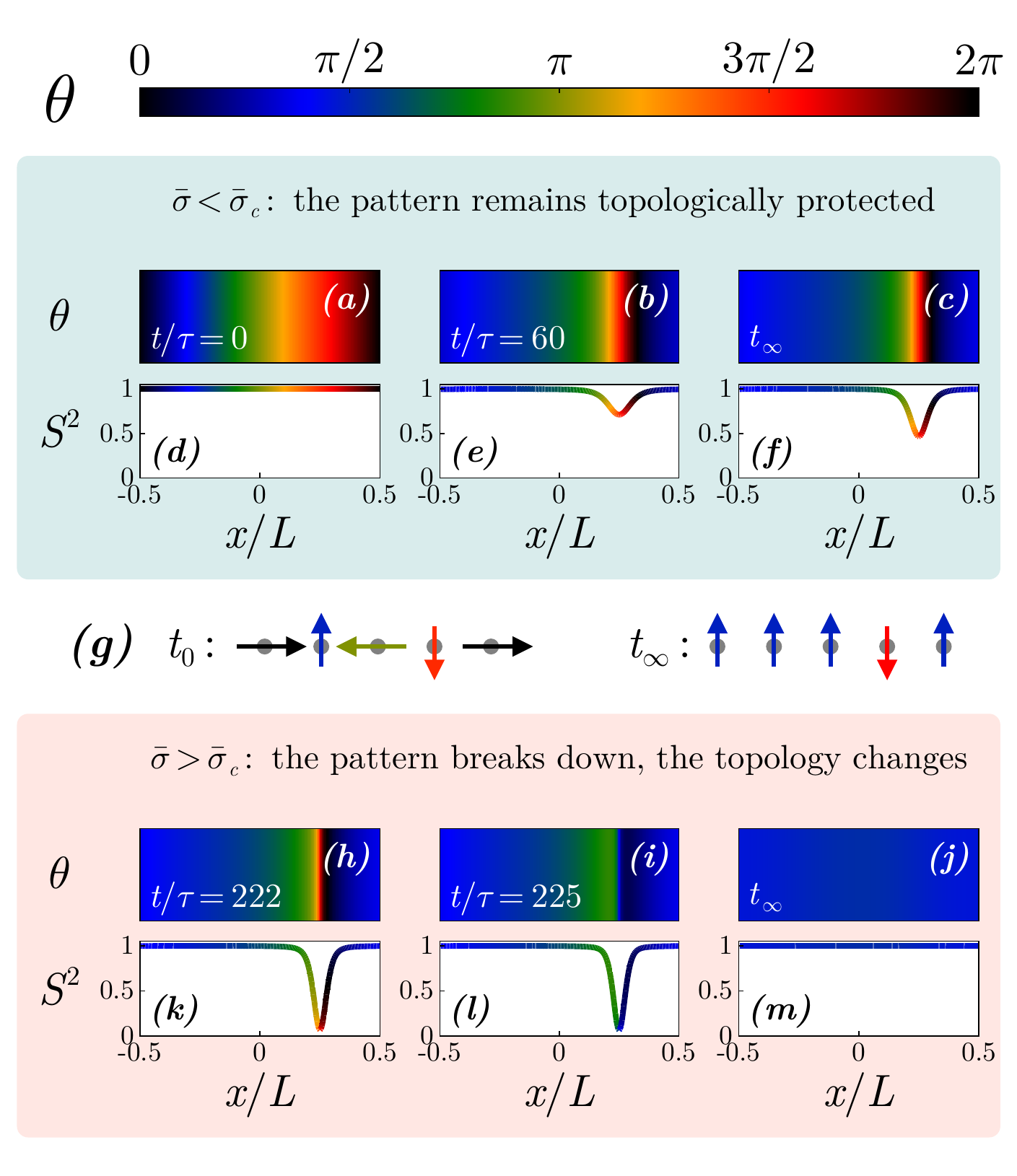}
    \caption{ Evolution of a topologically protected state over time, for two different values of $\bar{\sigma}$.   We show partial rectangular windows of the system, for $\alpha =100$. In panels (a-f), $\bar{\sigma}=3.9$ is below the critical value $ \bar{\sigma}_c$ (see explanation in the main text). 
    The orientation field $\theta$ is plotted in panels \textbf{(a-c)} for $t/\tau = 0, 60, 90$ (already in the steady state), respectively. The corresponding magnitude field $S^2$ is plotted in panels \textbf{(d-f)} for the same times. 
    In panels (h-m), $\bar{\sigma}= 4> \bar{\sigma}_c$ . $\theta$ is plotted in panels \textbf{(h-j)} for $t/\tau = 222, 225, 230$ (already in the steady state). $S^2$ is plotted in panels \textbf{(k-m)} for the same times. 
    In panel \textbf{(g)}, we sketch the mechanism that reshapes the spinwave.
  }
    \label{fig:snapshots_concentration}
\end{figure}

\begin{figure*}
    \centering
    \includegraphics[width=0.99\linewidth]{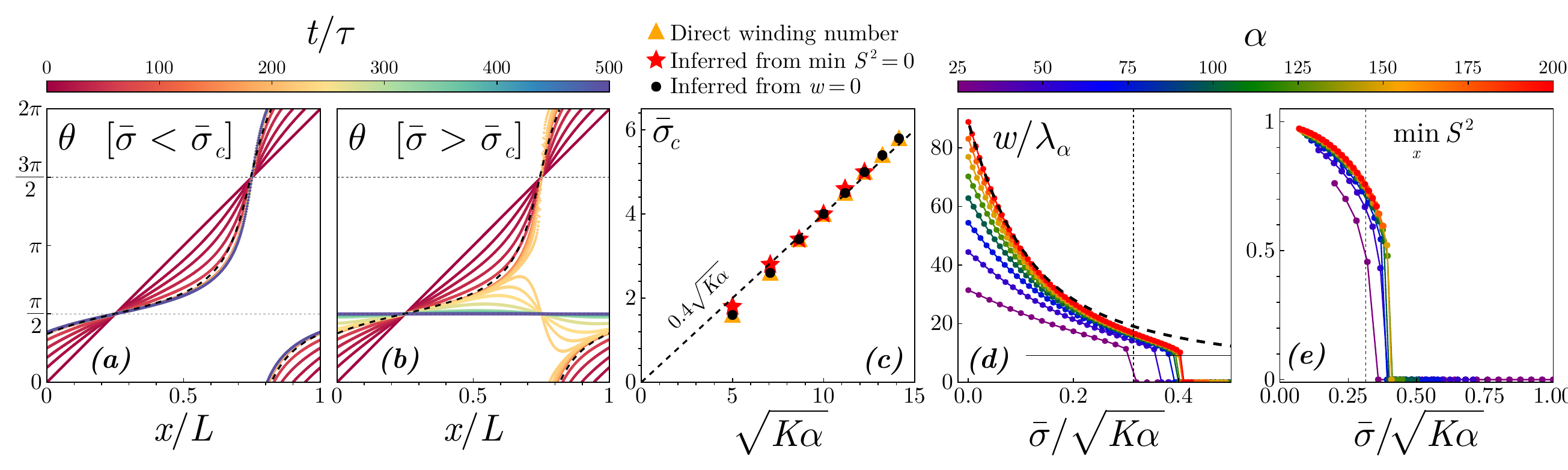}
    \caption{
   \textbf{(a)} Time evolution of the profile $\theta(x)$ for $\alpha=100, \bar{\sigma} =3.9$. The different lines and colors correspond to different times, from red ($t=0$, $\theta(x) = 2\pi\,x/L$) to blue (steady state). The dash line corresponds to Eq.~(\ref{eq:profile_derivation}).
    \textbf{(b)} Same as in panel (a) but for $\bar{\sigma} =4$. 
    The topology of $\theta(x)$ changes at $t/\tau=224$. 
\ylann{    \textbf{(c)} Critical non-reciprocity strength at from which the topology changes, computed using 3 different methods. }
   \textbf{(d)}     The steady width $w/\lambda_\alpha$ as a function of $\bar{\sigma}/\sqrt{K\alpha}$, for different $\alpha$ in different colors. The solid horizontal line is  $w = 10 \lambda_\alpha$. The dashed vertical line in (d) and (e) are $\bar \sigma / \sqrt{K\alpha}=\pi/10$, see main text. 
    \textbf{(e)} Value of the minimum of $S^2$ over the steady profile, against $\bar{\sigma} / \sqrt{K\alpha}$ , for different $\alpha$. A lower minimal value corresponds to a more compressed state.   
  }
    \label{fig:concentration1}
\end{figure*}

While Eq.~(\ref{eq:spinwave}) is a steady state solution of equilibrium $O(2)$ model, that is no longer the case when non-reciprocity is introduced in Eq.~(\ref{eq:main_eq}). Figure~\ref{fig:snapshots_concentration} illustrates the two possible scenarios for the evolution of the initial configuration given by Eq.~(\ref{eq:spinwave}) and illustrated in Fig.~\ref{fig:snapshots_concentration}(a). In both scenarios, the linear profile is initially deformed: gradients in the phase field get compressed around $\theta = 3\pi/2$ and stretched around $\theta = \pi/2$. Similar to an equilibrium $O(2)$ model, the magnitude of $S$ drops in the compressed regions to compensate for the increased energetic cost of the gradients in $\theta$, see Fig.~\ref{fig:snapshots_concentration}(d-f). {If the non-reciprocity is sufficiently low, $\bar{\sigma}<\bar{\sigma}_c$, this compressed spinwave is a stable configuration and the winding number remains equal to 1. Above a critical non-reciprocity $\bar{\sigma}_c$ \ylann{that we determine later}, the order parameter locally drops to zero. This allows the spins to reorient in a discontinuous fashion, breaking the topological protection of the spinwave and allowing it to dissipate, see Fig.~\ref{fig:snapshots_concentration}(h-m)}.

The initial evolution of the profile is easily understood from the agent-based perspective of the non-reciprocity. 
Let us sketch the initial profile at $t_0 =0 $ as in Fig.~\ref{fig:snapshots_concentration}(g, left). {Each spin tends to align with the spin it is pointing towards. The spins pointing directly up or down (blue and red), are stable since they point toward spins they are already aligned with. All other spins (black and green) point towards the blue arrow and away from the red arrow, thus tend to align vertically.}
The profile therefore tends to the one in Fig.~\ref{fig:snapshots_concentration}(g, right) as $t\to t_\infty$ and finally reaches a steady state where the gradients powered by non-reciprocity are compensated by the restoring elastic force.

{The state described by Eq.~(\ref{eq:spinwave}) is invariant in the $y$ direction so it is useful once again to consider the system in $1d$. Figure~\ref{fig:concentration1}(a) shows the evolution of the orientation of the spins from an initially linear configuration for $\bar{\sigma}<\bar{\sigma}_c$. In the long time limit it approaches a stable non-linear profile and the winding number {of the orientation field $\theta$ }is preserved. When non-reciprocity is increased beyond $\bar{\sigma}_c$, the non-linear configuration is no longer stable and the orientation field $\theta$ undergoes a topological change and the winding number goes to zero, see blue profiles in Fig.~\ref{fig:concentration1}(b).} \ylann{We define $\bar{\sigma}_c$ as the smallest $\bar{\sigma}$ that exhibits such a change in the winding number, and we plot it in Fig.~\ref{fig:concentration1}(c) with triangles. Dimensional analysis indicates that $\bar{\sigma}$ has units of $\sqrt{K\alpha}$, consistent with our fit $\bar{\sigma}_c = 0.4 \sqrt{K\alpha}$, see dash line in Fig.~\ref{fig:concentration1}(c). }

{To describe the stable non-linear profile when $\bar{\sigma}<\bar{\sigma}_c$, we assume the scalar order parameter $S=1$ and the dynamics in $1d$ can be described by Eq.~(\ref{eq:main_eq_1dof_1d}). The steady state solution of Eq.~(\ref{eq:main_eq_1dof_1d}) can be found explicitly through a change of variable, see SM~\cite{SM}, section I for details. For a system size of $L=2\pi$ the steady state profile is given by}
\begin{equation}
    \theta(x) = 2 \arctan \left\{\frac{1}{C}\, \left[ \frac{\bar{\sigma}}{K} +  \tan \Big( \frac{\Delta}{2} (x-x_0)\Big)\right]\right\} \ ,
    \label{eq:profile_derivation}
\end{equation}
with $\Delta = 2\pi/L$, $C^2 = \Delta^2+(\bar \sigma/K)^2 $ and $x_0 = 0.17\,L$ a constant of integration fitted to the data. 
We plot Eq.~(\ref{eq:profile_derivation}) as a dashed line in Fig.~\ref{fig:concentration1}(a) and we see that it is a good fit for simulated data. The only region where the prediction deviates from the data is the region $0.6<x/L<0.8$, where the drop in the order parameter $S$, see Fig.~\ref{fig:snapshots_concentration}(e,f), violates the $S=1$ assumption of Eq.~(\ref{eq:profile_derivation}). 

To follow the deformation of the spinwave, we define the steady state width $w$ of the compressed region as 
\begin{equation}
    w=\frac{2\pi}{ \partial_x\theta|_{\theta=3\pi/2}} \ . 
    \label{eq:w}
\end{equation}
The gradient is evaluated at the fixed point $\theta = 3\pi/2$, at which $\boldsymbol{S}$ points directly down and does not reorientate. 
For the initial linear profile Eq.~(\ref{eq:spinwave}), one obtains $w=L$ which then reduces as the gradients become condensed. Figure~\ref{fig:concentration1}(d) shows that the width of the stable profile decreases as $\bar{\sigma}$ is increased. For $\bar{\sigma} < \bar{\sigma}_{c}$, the stable width is the result of the competition between the non-reciprocal and the elastic forces. This can be obtained by substituting Eq.~(\ref{eq:profile_derivation}) into Eq.~(\ref{eq:w}) (see SM~\cite{SM}, section I) giving
\begin{equation}
 w = \frac{2\pi\,\lambda_{\bar{\sigma}}}{1+\sqrt{1 +\left(\frac{2\pi}{L}\lambda_{\bar{\sigma}} \right)^2 }}
 \label{eq:w_solution}
\end{equation}
where $\lambda_{\bar{\sigma}} = K/\bar{\sigma}$ is the length at which elastic and non-reciprocal forces balance. 
Equation~(\ref{eq:w_solution}) is shown in dashed lines (for $\alpha = 200$) in Fig.~\ref{fig:concentration1}(d) and predicts the numerical data very well.
If $\bar \sigma \to 0$, then $\lambda_{\bar{\sigma}} \to \infty$ and Eq.~(\ref{eq:w_solution}) gives $w\to L$, as expected. 
\ylann{For large non-reciprocity, $w \sim \pi \lambda_{\bar{\sigma}} \to 0$. Defining the critical non-reciprocity based on the $w=0$ criterion gives results similar to the original definition based on the winding number, see black circles in Fig.~\ref{fig:concentration1}(c).
As $\bar \sigma$ increases, $w$ eventually becomes comparable to $\lambda_\alpha$, the length scale at which distortions drive a local reduction in order parameter $S$, see Fig.~\ref{fig:snapshots_concentration}(k-l). 
We report the minimum value of $S^2$ against $\bar \sigma$ in Fig.~\ref{fig:concentration1}(e). The observed drop to 0 explains the breakdown dynamics.   
}

\begin{figure}
    \centering
    \includegraphics[width=0.99\linewidth]{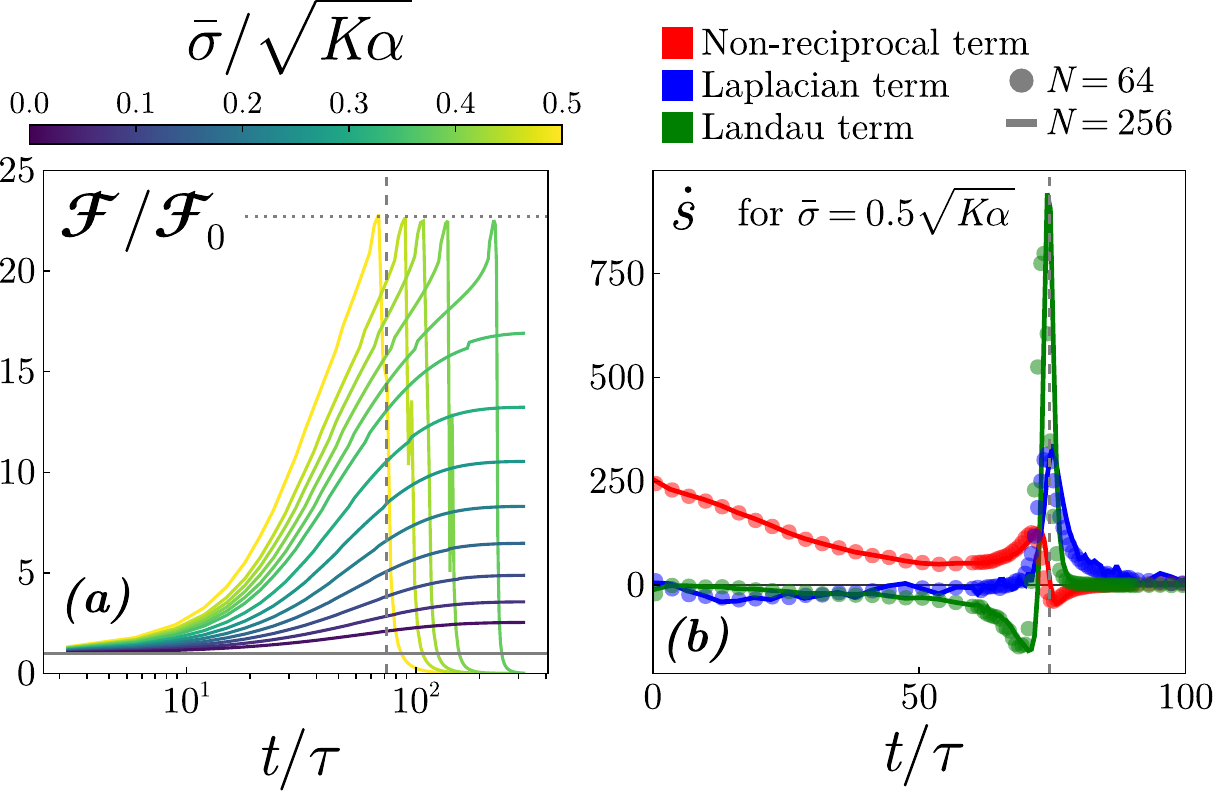}    
    \caption{
\ylann{    \textbf{(a)} Time evolution of the free energy $\mathcal{F}$, normalized by its initial value $\mathcal{F}_0 = \pi$, for $\alpha = 100$ and different $\bar{\sigma}$ (different colors). The dotted line at the top is to guide the eye to the common maximum of the curves ($\mathcal{F}_{\text{max}}  / \mathcal{F}_0 = 22.7$). In both panels, the vertical dashed line corresponds to the breakdown of the topological protection, measured with the winding number. 
        \textbf{(b)} The three contributions to entropy production rate $\dot s(t)$, each corresponding to one term in Eq.~(\ref{eq:eom}). 
       } }
    \label{fig:concentration2}
\end{figure}


Indeed, in the region where $S \approx 0$, the equation of motion Eq.~(\ref{eq:main_eq}) becomes largely independent of the orientation field $\theta$, thus allowing drastic reorientation, cf. the change from red to blue ($\theta = 3\pi/2 \to \theta = \pi/2$) between Fig.~\ref{fig:snapshots_concentration}(h) and \ref{fig:snapshots_concentration}(i). The corresponding $\theta$ profiles are plotted in Fig.~\ref{fig:concentration1}(b).  Such apparent change in topology is possible because of the relaxation of the spherical constraint: the amplitude of the field can locally vanish to allow its orientation to relax to its ground state.   
This is why the derivation of SM~\cite{SM}, section I, giving Eq.~(\ref{eq:profile_derivation}), cannot account for it, see dashes in Fig.~\ref{fig:concentration1}(b).
After the spinwave breaks down, the winding number is zero and the profile flattens to the steady state value $\theta = \pi/2$, as indicated in blue in Fig.~\ref{fig:concentration1}(b) or illustrated in Fig.~\ref{fig:snapshots_concentration}(j). At this point, the field $\theta$ is uniform, and $S$ tends to 1 everywhere, see Fig.~\ref{fig:snapshots_concentration}(m), thereby reaching the ground state of the equilibrium system. 

\ylann{We finally report energy and entropy production rate over the course of the compression and breaking of the spinwave state.}
The time evolution of the free energy $\mathcal{F}$, defined in Eq.~(\ref{eq:free_energy})  and normalized by its initial value {$\mathcal{F}_0 = 2K\pi^2/L$, }is plotted in Fig.~\ref{fig:concentration2}(a), for different $\bar{\sigma}$, increasing from blue to yellow. Active forces~$\sim\bar \sigma$ inject energy into the system. Interestingly, above $\bar{\sigma}_{c}$, this system provides a good example where non-reciprocity helps a system escaping a local minimum of the elastic energy $\mathcal{F}_0$ and reaching its {equilibrium, homogeneous ground state $(\mathcal{F}=0)$ where excitations travel as described in section 2.}
\ylann{We complement the measure of the free energy by the entropy production rate $\dot s$. Following \cite{kreienkamp2026entropy, cocconi2020entropy}, for a given vectorial force $\textbf{f}$, we defined it as 
\begin{equation}
    \dot s(t) = \frac{\langle \mathbf{f} \circ d\textbf{S}/dt \rangle}{k_B\, T}
\end{equation}
where $\textbf{S}$ is the orientational vector field, the brackets denote average over realisation and space and $\circ$ is a dot product understood in the Stratonovich convention \footnote{The Stratonovich convention implies to evaluate the force $\textbf{f}$ at time $t+dt/2$. Concretely, to compute $\dot s(t)$, we first run a standard Euler-Maruyama step, then compute   $d\textbf{S}/dt = (\textbf{S}(t+dt) - \textbf{S}(t))/dt$, define mid-point quantities as $\textbf{S}(t+dt/2) = (\textbf{S}(t+dt) + \textbf{S}(t))/2$ and evaluate the force $\textbf{f}$ at this mid-point.}. Defining entropy production requires a non-zero temperature $T$, so we manually add a small but finite $T=0.05$ to Eq.~(\ref{eq:eom}). 
We decompose $\dot s$ for the three forces entering Eq.~(\ref{eq:eom}), namely the Laplacian, the Landau term and the non-reciprocal curl term, and plot these three contributions in Fig.~\ref{fig:concentration2}(b), for $\bar \sigma = 0.5 \sqrt{K\alpha}$ to facilitate the comparison with the yellowmost curve of Fig.~\ref{fig:concentration2}(a). We have checked that varying the level of discretisation does not change the dynamics (circles for $N=64$, line for $N=256$). 
During the compression and before the breakdown (for $t/\tau < 50$), the dominant contribution is that of the non-reciprocal curl (in red). Just before the breakdown signalled by vertical dashed lines, the green Landau term contribution drops, reflecting the drop in $S$ from 1 to 0. The increase of the Laplacian term (in blue) reflects the drastic reorganisation of the orientation $\theta$, thereby increasing its gradients. Once the winding number has changed to 0, the magnitude very rapidly relaxes back to $S=1$, hence the dominant entropy production peak. Consistent with the observations of Fig.~  \ref{fig:snapshots_concentration}(h-m), the orientational field relaxes more slowly to a uniform field.
}

We have therefore shown that the topologically protected pattern Eq.~(\ref{eq:spinwave}) is unstable to non-reciprocal forces. Small values of non-reciprocity slightly deform the pattern, moderate values compress it to a narrow region of space and larger activity can even change the topology of the configuration and help the system to recover the ground-state of the equilibrium $\mathcal{O}(2)$ model, a scenario impossible without any active force. 
{We can extend these results to other values of the winding number, $k$. As $k$ is increased, the initial condition remains periodic over a length $L/k$, and features $2k$ stable points at which $\theta = \pi/{2}$ modulo $\pi$, as in the $k=1$ case. The $\theta$ profile between the fixed points is given by Eq.~(\ref{eq:profile_derivation}) scaled appropriately. A change in the sign of $k$ is equivalent to an inversion of the $y$ axis, otherwise all results remain the same.}

\section{Conclusions and perspectives}

\ylann{The recent discoveries on active matter called for a better understanding of dynamical phenomena taking place on non-reciprocal (NR) systems.
To characterise in depth unidirectional propagation of a perturbation, a generic feature found in many non-reciprocal systems, we introduced and studied a hydrodynamic description for the NR $2d$ XY lattice model. As most results can be derived analytically, it could become a canonical framework to comprehend vision-cone based non-reciprocity. 
\newline

Non-reciprocity introduces an additional active term to the equilibrium $O(2)$ model which responds to the curl of the vector field. This term bridges our model with the Toner-Tu model for polar flocks at constant density, highlighting the equivalence between non-reciprocity and activity. 
We have shown that propagation direction, speed and duration depend on the background medium, on the degree of non-reciprocity and on the perturbation itself. This behavior can be understood by mapping the equation of motion of a perturbation to a generalized Burgers equation. We have then applied this analysis to explain the propagation of localized perturbations in $2d$. 
We have also demonstrated that non-reciprocity breaks the spatial symmetry of global perturbations in which the orientation field has a non-zero winding number around the periodic boundaries. When non-reciprocity is sufficiently high, it is able to overcome the topological protection of the initial configuration and allows the system to reach the aligned state.
Altogether, our work paves the way to a comprehensive framework to describe how perturbations propagate in NR media and could help revisiting known results on wave propagation in NR (meta-) materials \cite{veenstra2025nonreciprocal, al2025non} or in living crowds such as Mexican waves in stadiums \cite{helbing2002mexicanwave}. 

\ylann{We have chosen to work at zero temperature to isolate the core impact of non-reciprocity. 
The impact of fluctuations (thermal or not) on this scenario thus remains elusive and would benefit a dedicated study, to address propagation phenomena at the micro scale, such as in beating cilia carpets \cite{hickey2023nonreciprocal}. }
Independently, studying a similar system conservation law (with conserved global magnetization for instance) to mitigate spreading and go beyond transient propagation could, 
combined with a refined control of the perturbation trajectory,
make a relevant and promising future perspective. 
\newline 
}


\paragraph{Acknowledgements}
D.L. thanks Ananyo Maitra, Sriram Ramaswamy and Sarah Loos for insightful discussions. 
Y.R. warmly thanks Ludovic Dumoulin for a game-changer introduction to GPU programming in Julia. 
Y.R. thanks Sriram Ramaswamy and Ricard Alert for insightful discussions. 

D.L. acknowledges support by grant NSF PHY-2309135 to the Kavli
Institute for Theoretical Physics  and MCIU/AEI for financial support under
grant agreement PID2022-140407NB-C22. 
D.L. and I.P. acknowledge DURSI for financial support under Project No. 2021SGR-673. 
I.P.  and Y.R. acknowledge support from Ministerio de Ciencia, Innovaci\'on y
Universidades MCIU/AEI/FEDER for financial support under
grant agreement PID2024-156516NB-100 AEI/FEDER-EU.
I.P. acknowledges Generalitat de Catalunya for financial support under Program Icrea Acad\`emia.
D.J.G.P. acknowledges Swiss National Science Foundation for financial support under SNSF Starting Grant TMSGI2\_211367.  

\bibliography{biblio}

\appendix
\onecolumngrid
\newpage

\begin{center}
    \textsc{Appendices}
\end{center}
\ \newline

\textbf{\textit{Table of content}} \ \\ 
\newline
Movies: \\
\textbf{Appendix A} : Rewriting the results of \cite{dopierala2025inescapable} \\
\textbf{Appendix B} :  Emission cones and reception cones  \\
\textbf{Appendix C} : Natural scales and implementation details  \\
\textbf{Appendix D} :  Conserved area under the curve  \\
\textbf{Appendix E} :  Dimensional analysis, spreading and $\beta$ exponent  \\
\textbf{Appendix F} :  Impact of $\theta_0$ on $1d$ propagation  \\
\textbf{Appendix G }:  Decay rates of annihilating peaks for the diffusion equation \\
\textbf{Appendix H} :  Quartic excitation  \\
\textbf{Appendix I} :  Derivation of the stationary solution of the TPP \\

\section{Movies}
We provide some movies to help the reader understand the system dynamics. \\
They can be found on \href{https://www.youtube.com/watch?v=NmF8G89lxNU&list=PLDLpeAHw9KhfmUEzb44rD6Ik0m4re7XJ0&pp=sAgC}{this Youtube channel}. 
\newline

Like in the main text, all the simulations are performed with periodic boundary conditions and for $\gamma = 1, K=1, \alpha = 100, L=2\pi$.  For $1d/2d$ gaussian perturbations, the initial standard deviation is $w_0/L = 0.05$. Here follows a description of each movie. \\
\textbf{Movie 1}: 1$d$ gaussian perturbation, small non-reciprocity $\bar \sigma$.  The background orientation $\theta_0=0$ so the perturbation travels to the left. The excitation travels over a distance $\sim 1.5L$. \\
\textbf{Movie 2}: 1$d$ gaussian perturbation, large non-reciprocity $\bar \sigma$ to see the long term propagation of the perturbation (also to the left because $\theta_0=0$). The excitation travels over a distance $\sim 15L$. \\
\textbf{{Movie 3}}: 1$d$ gaussian perturbation. In contrast with Movies 1 and 2, here the background orientation $\theta_0=\pi/4$ and the initial amplitude of the perturbation $\delta_0=\pi/2$. As such, at very short times, the perturbation peak travels to the right, whereas the lower part of the perturbation travels to the left.\\
\textbf{{Movie 4}}: 1$d$ gaussian perturbation.  The background orientation $\theta_0=\pi/2$ is perpendicular to the only possible direction of travel (along $x$, because invariance in the $y$ direction). This medium cannot sustain a long term propagation. The excitation travels over a distance $\sim L/2$ before vanishing. \\
\textbf{{Movie 5}}: 2$d$ gaussian perturbation. \\
\textbf{Movie 6}: 2$d$ perturbation odd in the $x$ direction: $\delta(x,y,t=0) = \delta_0\sin({\color{red}{x}}/w_0) \, \exp\left(-\frac{x^2 + y^2}{2\,w_0^2} \right)$. \\
Movie 6a, $\delta_0=+\pi/2$, Movie 6b, $\delta_0=-\pi/2$. \\
\textbf{{Movie 7}}: 2$d$ perturbation odd in the $y$ direction: $\delta(x,y,t=0) = \delta_0\sin({\color{red}{y}}/w_0) \, \exp\left(-\frac{x^2 + y^2}{2\,w_0^2} \right)$. \\
Movie 7a, $\delta_0=+\pi/2$, Movie 7b, $\delta_0=-\pi/2$. \\
\textbf{{Movie 8}}: $1d$ topological protected pattern. Left:  $\sigma < \, \sigma_c$ so the pattern gets compressed but the topology is conserved. Right:  $\sigma > \, \sigma_c$ , the pattern gets so compressed that there is a topological breakdown.  

\section{Appendix A. Rewriting the results of \cite{dopierala2025inescapable} }\label{sec:rewriting_chate}

In this section, we explicit the simple algebraic rewriting of the Equation (2) of \cite{dopierala2025inescapable}, in our notation (their $J_0$ is our $J$, their $J_1$ is our $\sigma$). We show that the Toner-Tu like equations, with the coefficient derived in \cite{dopierala2025inescapable}, can almost be rewritten in terms of a curl term and a functional derivative term.   
\begin{align}
\lambda_1 &= -\tfrac{\sigma}{2}\left(-3 + \tfrac{J}{T}\right), \\[6pt]
\lambda_2 &= -\tfrac{\sigma}{2}\left(1 - \tfrac{J}{T}\right), \\[6pt]
\lambda_3 &= -\tfrac{\sigma}{4}\left(1 + \tfrac{J}{T}\right).
\end{align}

One can show that 
\[
\lambda_3 = -\tfrac{1}{2}\lambda_1 - \lambda_2,
\]
so
\begin{align}
& \ \lambda_1 (\boldsymbol{S}\cdot\nabla)\boldsymbol{S} 
  + \lambda_2 (\nabla\cdot\boldsymbol{S})\boldsymbol{S} 
  + \lambda_3 \nabla |\boldsymbol{S}|^2 \\[6pt]
= &\ \lambda_1 \Big[ (\boldsymbol{S}\cdot\nabla)\boldsymbol{S} 
     - \tfrac{1}{2}\nabla|\boldsymbol{S}|^2 \Big] 
   + \lambda_2 \Big[ (\nabla\cdot\boldsymbol{S})\boldsymbol{S} 
     - \nabla|\boldsymbol{S}|^2 \Big] \\[6pt]
= &\ \lambda_1 (\nabla\times\boldsymbol{S})\times\boldsymbol{S} 
   + \lambda_2 \Big[ (\nabla\cdot\boldsymbol{S})\boldsymbol{S} 
     - \nabla|\boldsymbol{S}|^2 \Big] \\[6pt]
=& \ \lambda_1 (\nabla\times\boldsymbol{S})\times\boldsymbol{S} 
   + \frac{\lambda_2 }{2}\left\{\,\frac{\delta}{\delta \boldsymbol{S}}
      \Big[ (\nabla\cdot\boldsymbol{S})|\boldsymbol{S}|^2 \Big]  - \nabla|\boldsymbol{S}|^2\right\} \ .
\end{align}

We thank Sriram Ramaswamy for pointing out that a free energy of the form $(\nabla\cdot\boldsymbol{S})|\boldsymbol{S}|^2 $ could give rise to the $\lambda_2$ and $\lambda_3$ terms. \\

In \cite{dopierala2025inescapable}, the Lagrange term is 
\begin{equation}
    (\alpha - \gamma|\boldsymbol{S}|^2)\boldsymbol{S} \ .
\end{equation}
To compare with our framework $(1 - |\boldsymbol{S}|^2)\boldsymbol{S}$,  which penalizes magnitude deviations from \textit{unity}, one simply applies the rescaling 
\begin{equation}
    \boldsymbol{S} \to \boldsymbol{S} \, \sqrt{\alpha/\gamma} \ .
\end{equation}
Note that this rescales the $\lambda$ coefficients: 
\begin{equation}
    \lambda \to \lambda \sqrt{\alpha/\gamma} \ . 
\end{equation}

\section{Appendix B. Emission cones and reception cones }

In a fully reciprocal system, each particle emits and receives information isotropically, independently of the orientation of the emitter or of the receiver. 

When equipped with a \textit{reception cone}, a particle $i$ emits information (=its orientation $\theta_i$) isotropically but the reception of the neighbors' information $\theta_j$ is weighted by a kernel depending on the orientation of the receiver: the reception kernel $g$ depends on $\text{Angle}(\boldsymbol{\hat S}_{\color{red}{i}}\, ,   \boldsymbol{u}_{ij})$. In other words, you will receive a higher signal if you look at your neighbor, but the amplitude of this information does not explicitly depend on the value $\theta_j$. An intuitive realization of such a system is precisely the well-known \textit{vision cone} in the animal kingdom. 

However, one can think of the reversed case, ie. \textit{emission cones}, where each agent perceives information isotropically but emits it in a preferential direction. This is for instance the case for directed sound emission of animals. There, the emission is angularly weighted by the orientation of the emitter and the kernel $g$ depends on $\text{Angle}(\boldsymbol{\hat S}_{\color{red}{j}}\, ,   \boldsymbol{u}_{ij})$. You will receive a higher signal if your neighbor looks at you, but this signal does not depend on your own orientation. 
\newline

This motivates the study of the following equation of motion, which describes a system of agents with emission cones (note the red index $j$ instead of the index $i$ for the reception cones, and the change of the sign of $\sigma$)
\begin{equation}
    \dot{\theta}_i  =\sum_j \Big[1-\sigma \cos \left(\theta_{\color{red}{\textbf{j}}}-u_{i j}\right)\Big] \sin \left(\theta_j-\theta_i\right) + \sqrt{2T} \eta_i(t) \ .
\end{equation}
Working on a square lattice at $T=0$, one can explicitly write the sum on the 4 nearest neighbors to obtain
\begin{equation}
\begin{aligned}
\dot{\theta}_i & =\left(1-\sigma \cos \theta_1\right) \sin \left(\theta_1-\theta_i\right) \\
& +\left(1-\sigma \cos \left(\theta_2-{\pi}/{2}\right)\right) \sin \left(\theta_2-\theta_i\right) \\
& +\left(1-\sigma \cos \left(\theta_3-\pi\right)\right) \sin \left(\theta_3-\theta_i\right) \\
& \left.+1-\sigma \cos \left(\theta_4+\pi / 2\right)\right) \sin \left(\theta_4-\theta_i\right) \\
\dot{\theta}_i & =\left(1-\sigma \cos \theta_1\right) \sin \left(\theta_1-\theta_i\right) \\
& +\left(1-\sigma \sin \theta_2\right) \sin \left(\theta_2-\theta_i\right) \\
& +\left(1+\sigma \cos \theta_3\right) \sin \left(\theta_3-\theta_i\right) \\
& +\left(1+\sigma \sin \theta_4\right) \sin \left(\theta_4-\theta_i\right)
\end{aligned}\\
\end{equation}
\begin{equation}
\begin{aligned}
& \dot{\theta}_i=\sum_j \sin \left(\theta_j-\theta_i\right) \\
& -\sigma\left\{\cos \theta_1 \sin \left(\theta_1-\theta_i\right)-\cos \theta_3 \sin \left(\theta_3-\theta_i\right) +\sin \theta_2 \sin \left(\theta_2-\theta_i\right)-\sin \theta_4 \sin \left(\theta_4-\theta_i\right)\right\}
\end{aligned}
\end{equation}

For the non-reciprocal part in $\sigma$, one can approximate, to first order in the Taylor expansion ($a$ is the lattice spacing)
\begin{equation*}
\begin{aligned}
& \theta_1=\theta_i+a \frac{\partial \theta}{\partial x} \equiv \theta_i+\theta_x \\
& \theta_3=\theta_i-\theta_x \\
& \theta_2=\theta_i+a \frac{\partial \theta}{\partial y} \equiv \theta_i+\theta_y \\
& \theta_4=\theta_i-a \frac{\partial \theta_y}{\partial y}=\theta_i-\theta_y
\end{aligned}
\end{equation*}
one thus obtains
\begin{equation*}
\begin{aligned}
& \quad \cos \left(\theta+\theta_x\right) \sin \theta_x+\cos \left(\theta-\theta_x\right) \sin \theta_x \\
& +\sin \left(\theta+\theta_y\right) \sin \theta_y-\cos \left(\theta-\theta_y\right) \sin \theta_y
\end{aligned}
\end{equation*}

If one assumes small gradients, one can approximate $\cos \alpha \approx 1$ and $\sin \alpha \approx \alpha$. One thus obtains
\begin{equation*}
2\left(\theta_x \cdot \cos \theta+\theta_y \sin \theta\right)=2(\nabla \times \hat{\textbf{S}})_z \ .
\end{equation*}

For the reciprocal part, one needs to go up to the second order in the Taylor expansion because the first orders are equal and opposite, leaving us with no reciprocal part and $\dot\theta = \mathcal{O}(\sigma)$ 
\begin{equation*}
\begin{aligned}
& \theta_1=\theta_i+a \frac{\partial \theta}{\partial x}+\frac{1}{2} a^2 \frac{\partial^2 \theta}{\partial x^2} \equiv \theta_i+\theta_x+\theta_{x x} \\
& \theta_3=\theta_i-a \frac{\partial \theta}{\partial x}+\frac{1}{2} a^2 \frac{\partial^2 \theta}{\partial x^2} \equiv \theta_i-\theta_x+\theta_{x x} \\
\end{aligned}
\end{equation*}
Which gives
\begin{equation*}
\sin \left(\theta_1-\theta_i\right)+\sin \left(\theta_3-\theta_i\right) = \sin \left(\theta_x+\theta_{x x}\right)+\sin \left(-\theta_x+\theta_{x x}\right) = 2 \sin \theta_{x x} \cos \theta_x
\end{equation*}

which, in the small gradients limit, is simply equal to $\frac{\partial^2 \theta}{\partial x^2}$. 
Following the same procedure for the $y$ axis ( with $\theta_2$ and $\theta_4$ ), one obtains the Laplacian $\Delta \theta=\frac{\partial^2 \theta}{\partial x^2}+ \frac{\partial^2 \theta}{\partial y^2}$
Finally, one obtains
\begin{equation*}
\dot \theta=a^2 \Delta \theta-2 a\, \sigma (\nabla \times \hat{\textbf{S}})_z \ .
\end{equation*}
Both microscopic models are different (up to a $\sigma \to -\sigma$ transformation, which we know can be absorbed by a $\theta\to \theta+\pi$ global rotation), yet they lead to the same hydrodynamic equation. In other words, at the continuum level, receiving information from ahead and emitting it isotropically is equivalent to emitting information to the back and receiving it isotropically. In both cases, information is advected by the local polarization field.
\newline

This observation might be useful in cases where it is experimentally simpler to design a setup (of interactions based on light, sound, chemicals, ...) where the emission, rather than the reception, is anisotropic. 
\newline

\section{Appendix C. Natural scales and implementation details}\label{sec:dimensional_analysis}
In this section, we extract the natural scales from Eq.~(\ref{eq:main_eq}) 
\begin{equation}
\gamma\dot{\boldsymbol{S}} = K\Delta\boldsymbol{S}
\!+ \bar{\sigma} (\nabla \times \boldsymbol{S})\times \boldsymbol{S} + \alpha (1-|\boldsymbol{S}|^2)\boldsymbol{S} \ .
\end{equation}
Because $\boldsymbol{S}$ is an orientation vector field, it has no units, so Eq.~(\ref{eq:main_eq}) is dimensionally equivalent to 
\begin{equation}
\frac{\gamma}{t}= \frac{K}{x^2} + \frac{\bar \sigma}{x} + \alpha\ ,
\end{equation}
where $t$ means time and $x$ means space. Each term has to have the same unit. 
From there, one deduces the only time scale of our problem: 
\begin{equation}
    \tau = \gamma / \alpha \ .
\end{equation}
One can construct 3 distinct length scales: 
\begin{equation}
 \lambda_\alpha  =    \sqrt{\frac{K}{\alpha}}\ , \  \frac{\bar \sigma}{\alpha} \text{ and } \lambda_{\bar \sigma}  =\frac{K}{\bar \sigma} \ .
\end{equation}

The length $ \lambda_\alpha$ is the length at which distortions drive a local reduction in order parameter $S$. 
The length $\lambda_{\bar \sigma}$ is the length at which elasticity and non-reciprocal forces balance.
Following, one can construct 3 speed scales: 
\begin{equation}
    \lambda_\alpha/\tau = \frac{\sqrt{K\alpha}}{\gamma}\ , \ \frac{\bar \sigma}{\gamma} \text{ and }  \lambda_{\bar \sigma}/\tau = \frac{K\alpha}{\bar \sigma \,\gamma} \ .
\end{equation}

The natural scale for diffusion coefficients is $K/\gamma$ and the natural scale for non-reciprocity is $\sqrt{K{\alpha}}$ . 
Note that, once space and time scales are fixed, the ratio $ \lambda_{\alpha}/  \lambda_{\bar \sigma} = \bar \sigma / \sqrt{K\alpha} $ is the only free adimensional parameter and represents the activity / non-reciprocity level in the system.  
\newline 

We integrate Eq.~(\ref{eq:main_eq})  with the Euler method with timestep $dt$. 
We approximate the gradients using finite difference methods, on a $N\times N$ square mesh, such that $dx = L/N$. 
We set $L=2\pi$, and use $\alpha = 100, K=1, \gamma=1$ unless specified otherwise. 
One requires $dx \ll \lambda_\alpha = \sqrt{K/\alpha} \sim 0.1$. 
Therefore, we choose $N= 256$. 
To correctly resolve the dynamics in time, the difference in angle in one timestep $\Delta \theta \, dt$ has to be small, under the (arbitrary) threshold $\pi/20$. 
In particular, the Laplacian (the most constraining) implies $d^2/dx^2 \,dt = N^2/L^2\,dt \le \pi/20$, implying $dt \le 10^{-4}$. We choose $dt = 10^{-5} =  10^{-5}\tau$.

\section{Appendix D. Conserved area under a perturbation }
We consider a perturbation $\delta(x,t)$ over a constant background with orientation $\theta_0$, thus $\theta(x,t) = \theta_0 + \delta(x,t)$. We define the mass $M$, ie. the area under the perturbation as
\begin{equation}
    M(t) = \int\limits_0^L \delta(x,t) dx \ .
\end{equation}
Here we show that $M$ is constant over time, by proving that $\dot M = 0 $ . 
\begin{align}
   \dot M &= \frac{\partial}{\partial t}\int\limits_0^L \delta \ dx \\
   \dot M &= \int\limits_0^L \frac{\partial \delta}{\partial t} \ dx 
\end{align}
We now substitute in Eq.~(\ref{eq:perturbation1_dynamics}) and define $\delta_{x} = \partial\delta/\partial x$ and $\delta_{xx} = \partial^2\delta/\partial x^2$. 
\begin{align}
   \gamma \dot M &= \int\limits_0^L K\delta_{xx} + \bar \sigma \cos (\theta )\delta_x \ dx \\
\gamma \dot M &= \int\limits_0^L K\delta_{xx} \ dx + \int\limits_0^L \bar \sigma \cos (\theta )\delta_x \ dx \\
\gamma \dot M &= K\int\limits_0^L \delta_{xx} \ dx + \bar \sigma\int\limits_0^L \frac{\partial}{\partial x}\sin \theta  \ dx \\
\gamma \dot M &=  \left. \phantom{\frac12}K\delta_{x} \right|_0^L \phantom{\frac12} +  \left.\phantom{\frac12}\bar \sigma \sin (\theta )\right|_0^L \\
\gamma \dot M &=0
\end{align}
The quantities evaluated between 0 and $L$ are identically zero due to periodic boundary conditions (PBC). Even without PBC, if one were to evaluate on an infinite interval (much larger than the width of the perturbation), the result would be the same. Indeed, since $\theta = \theta_0 + \delta$, with $\lim\limits_{|x|\to \infty} \delta(x) = 0$ and $\lim\limits_{|x|\to \infty} \delta_x(x) = 0$, the perturbation vanishes at infinity. 

The second integral vanishes because its integrand can be written as a derivative along $x$ of another function. This is not a coincidence. This follows from the integrand stemming from a rotational, which, when $\partial_y = 0$ as it is the case here, is by definition equal to $\partial_x (...)$. 
\newline 

Thus, $\dot M = 0$  and the mass under the curve is constant over time. 
\section{Appendix E.  Dimensional analysis, spreading and exponent $\beta$ }

We consider a generalized $1d$ Burgers equation, where we define $\delta_{x} = \partial\delta/\partial x$ and $\delta_{xx} = \partial^2\delta/\partial x^2$ : 
\begin{equation}
\partial_t \delta=\frac{K}{\gamma} \delta_{xx}+ \frac{\bar{\sigma} \, \sin \theta_0}{\gamma}\, \delta\,\delta_x + \frac{\bar{\sigma} \, \cos \theta_0}{\gamma}\, \delta^2\delta_{x} \ .
\end{equation}

We have shown in Appendix D that the mass under the perturbation $M$ is constant over time. 
If the profile has a typical height $h(t)$ and a typical width $w(t)$, then $M \sim h(t)\,w(t)$
which implies $h(t)\sim 1/{w(t)}$~. 
We thus seek a self-similar spreading law
\begin{equation}
w(t)\sim t^{\beta},
\qquad
h(t)\sim t^{-\beta},
\end{equation}
and use the scaling form
\begin{equation}
\delta(x,t)\sim t^{-\beta}F\!\left(\xi\right),
\qquad
\xi=\frac{x}{t^{\beta}}.
\end{equation}
We note $F'(\xi) = {\partial F(\xi)}/{\partial \xi}$ and $F''(\xi) = {\partial^2 F(\xi)}/{\partial \xi^2}$ . Note that $\xi_x = t^{-\beta}$, $\xi_{xx} = 0$ and $\xi_t \sim t^{-\beta-1}$. 
Under the above ansatz, derivatives scale as
\begin{equation}
    \delta_t = \frac{\partial \delta}{\partial t} = \left(t^{-\beta-1} F(\xi) + t^{-\beta} F'(\xi)\, \xi_t\right) \sim t^{-\beta-1} \text{ to leading order}
\end{equation}
\begin{equation}
    \delta_x = \frac{\partial \delta}{\partial x} = t^{-\beta}F'(\xi)\, \xi_x \sim t^{-2\beta}
\end{equation}
\begin{equation}
    \delta_{xx} = \frac{\partial \delta_x}{\partial x} = t^{-\beta}\left[ F''(\xi)\, (\xi_x)^2 + F'(\xi)\, \xi_{xx} \right] \sim t^{-3\beta}
\end{equation}

{Note here that functions of higher orders of $\xi$, i.e. $F(\xi^2)$ lead to higher order terms that decay faster, thus are discarded here. We use the above estimates to estimate the scale of each term in the generalized Burgers equation, arriving at:}
\begin{align}
    \delta_t&\sim t^{-\beta-1}\\
    \delta_{xx}&\sim t^{-3\beta}\\
    \delta\delta_{x}&\sim t^{-3\beta}\\
    \delta^2\delta_{x}&\sim t^{-4\beta}
\end{align}
{Thus at long times, terms that scale with $t^{-3\beta}$ dominate which implies}
\begin{align}
t^{-\beta-1}&\sim t^{-3\beta}\\
\implies \beta&=\frac12.
\end{align}

Therefore,
\begin{equation}
w(t)\sim t^{1/2},
\qquad
h(t)\sim t^{-1/2}.
\end{equation}

{However, at short times terms that scale with $t^{-4\beta}$ dominate.} Balancing $\delta_t$ with $\delta^2\delta_{x}$ gives
\begin{align}
    t^{-\beta-1}&\sim t^{-4\beta}\\
    \implies \beta&= \frac13.
\end{align}
Therefore,
\begin{equation}
w(t)\sim t^{1/3},
\qquad
h(t)\sim t^{-1/3}.
\end{equation}

\paragraph{Crossover time} \ \\

The exponents $\beta$ follow from scaling balances and are therefore universal within their respective regimes. We now address the crossover window from one pure scaling regime ($\beta=1/3$) to the other ($\beta=1/2$). We do so by comparing the amplitude of the $\delta^2\,\delta_x$ term, giving the $\beta=1/3$ scaling, to then amplitude of the two terms $\delta\,\delta_x$ and $\delta_{xx}$, giving the $\beta=1/2$ scaling. For now, let us leave $\delta \sim (t/\tau)^{-\beta}$ as is, without changing it for its numerical value. Recall that $\tau = \gamma /\alpha \ , \ \lambda_{\bar \sigma} = K/\bar \sigma$ and $\lambda_{\alpha}/\lambda_{\bar{\sigma}}   = \bar \sigma /\sqrt{K\alpha}$ .
\newline

The $\delta^2\,\delta_{x}$ and the $\delta\,\delta_{x}$ terms are thus comparable when 
\begin{align}
    \frac{\bar \sigma \, \cos \theta_0}{\gamma}\,\delta^2\,\delta_{x} & \sim \frac{\bar \sigma \, \sin \theta_0}{\gamma}\,\delta\,\delta_{x} \\
\delta\,  \cos \theta_0 &\sim \sin \theta_0  \\
\left( \frac{t}{\tau}\right)^{-\beta}\,  \cos \theta_0 &\sim \sin \theta_0  \\
\frac{t}{\tau} &\sim \tan^{1/\beta} \theta_0  \ .
\end{align}

The $\delta^2\,\delta_{x}$ and the $\delta_{xx}$ terms are comparable when
\begin{align}
     \frac{\bar \sigma \, \cos \theta_0}{\gamma}\,\delta^2\,\delta_{x} & \sim \frac{K}{\gamma}\,\delta_{xx} \\
          \bar \sigma \, \cos \theta_0\,\delta^2\,\delta_{x} & \sim K\,\delta_{xx} \\
          \bar \sigma \, \cos \theta_0\,\left( \frac{t}{\tau}\right)^{-4\beta} \frac{1}{\lambda_\alpha} & \sim K\, \left( \frac{t}{\tau}\right)^{-3\beta} \frac{1}{\lambda_\alpha^2} \\
                    \bar \sigma \, \cos \theta_0\,\lambda_\alpha & \sim K\, \left( \frac{t}{\tau}\right)^{\beta}  \\
                     \frac{\lambda_\alpha}{\lambda_{\bar{\sigma}}} \, \cos \theta_0 & \sim \left( \frac{t}{\tau}\right)^{\beta}  \\
                    \left( \frac{\bar \sigma}{\sqrt{K\alpha}} \, \cos \theta_0\right)^{1/\beta} & \sim  \frac{t}{\tau}  \ .
\end{align}
We define 
\begin{equation}
    \zeta \sim \min\Big \{\frac{1}{\tan \theta_0}, \frac{\bar{\sigma} \cos \theta_0 }{\sqrt{K\alpha}}\Big \} \ , 
\end{equation}
which can be controlled by changing the parameters of the model, in particular $\bar{\sigma}$ and $\theta_0$. \\
Replacing $\beta$ by $1/3$ gives the time scale $\tau\, \zeta^3$ , at from which the $\delta^2\,\delta_{x}$ term becomes subdominant. Replacing $\beta$ by $1/2$ gives the time scale $\tau\, \zeta^2$ at which one departs from the pure $\beta=1/2$ Burgers regime because the $\delta^2\,\delta_{x}$ term becomes dominant. \\
The crossover window therefore extends from  $\tau\, \zeta^2$ to $\tau\, \zeta^3$. During this crossover time, the amplitude of the perturbation cannot be expressed with a single power law $\delta_{\text{peak}} \sim t^{-\beta}$.

\section{Appendix F. Impact of $\theta_0$ on $1d$ propagation} \label{sec:profiles_different_theta0}
In the spirit of the Figure~\ref{fig:propagation}(a,b), we plot the profiles $\theta(x)$ of a $1d$ perturbation over time, from $t=0$ in red to $t_\text{max} = 0.2$ in blue, for different background orientations $\theta_0 = n\pi/8 \text{, with } n = 0, ..., 15$, as indicated above each panel. 

\begin{figure*}
    \centering
    \includegraphics[width=\linewidth]{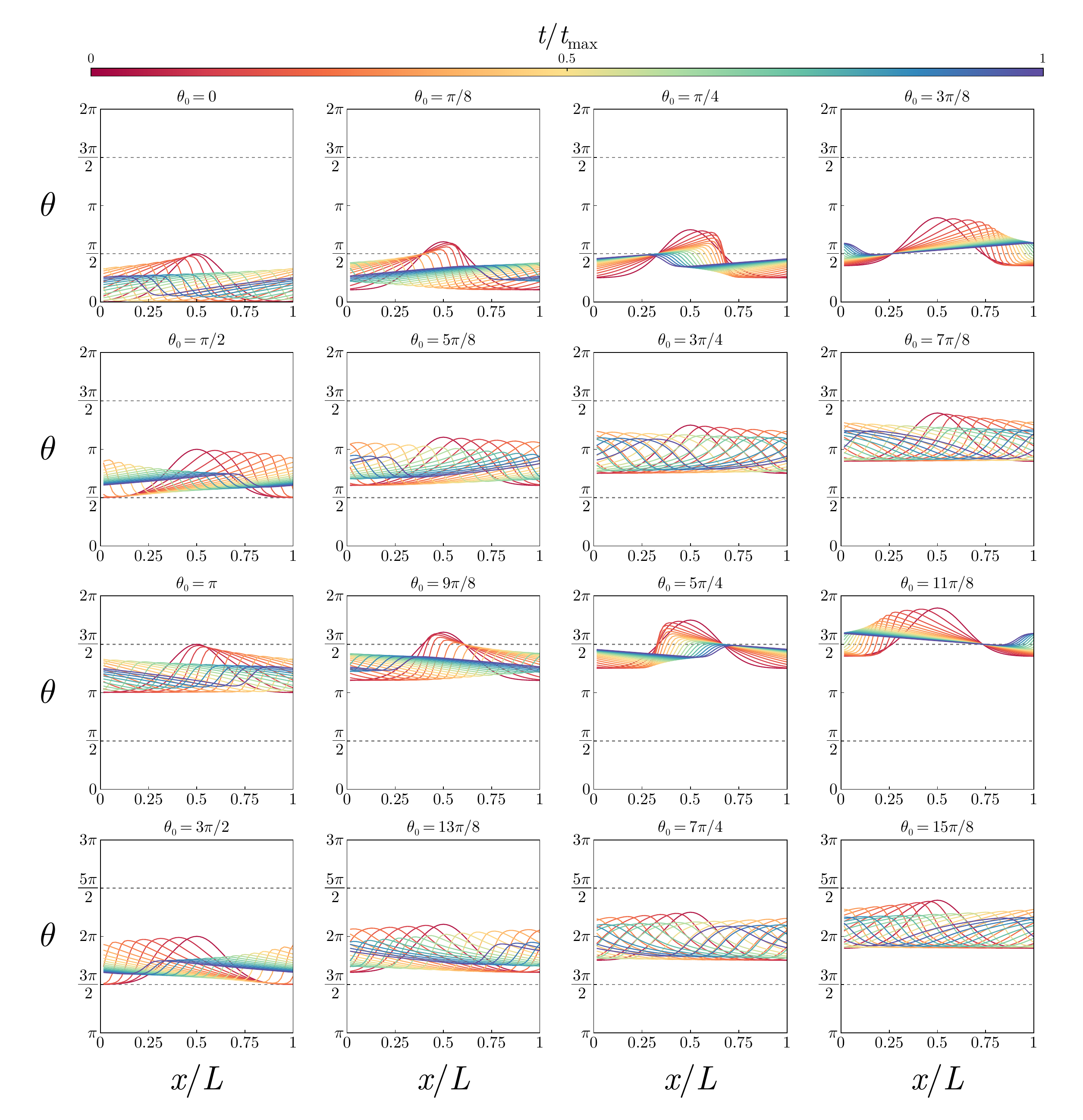}    
    \caption{
    Profiles $\theta(x)$ of a 1d perturbation over time, from $t=0$ in red to $t_\text{max} = 0.2$ in blue, for different background orientations $\theta_0 = n\pi/8 \text{, with } n = 0, ..., 15$, as indicated above each panel. \\ We plot horizontal dashed lines to mark the $\theta = \pi/2$ and $\theta = 3\pi/2$ lines, where the information flux changes sign and orientation. For $\pi/2<\theta<3\pi/2$, the profiles are advected to the right, for $\theta<\pi/2$ and $\theta>3\pi/2$, the profiles are advected to the left.  
Parameters : $L=64, \bar{\sigma} = 60, t_\text{max}=0.2, \alpha = 100 $}
    \label{fig:profiles_different_theta0}
\end{figure*}
\newpage

\section{Appendix G. Decay rates of annihilating peaks for the diffusion equation}

When non-reciprocity is removed ($\bar{\sigma}=0$) the $1d$ dynamics of a perturbation given in Eq.~(\ref{eq:perturbation1_dynamics}) reduce to a diffusion equation with diffusion constant $D=K/\gamma$. The solution to the diffusion equation in $1d$ at time $t$ for an initial condition $\delta_0(x)$ is given by 
\begin{equation}
    \delta(x,t) = \int_{-\infty}^\infty G(x-x',t)\delta_0(x')\mathrm{d}x'.
\end{equation}
Here $G$ is the Green's function of the diffusion equation which is a Gaussian given by
\begin{equation}
    G(x,t) = \frac{1}{\sqrt{4\pi Dt}}\exp\left(-\frac{x^2}{4Dt}\right)
\end{equation}
where $D = K/\gamma$ is the diffusion constant.

In the long time limit, the width of $G$ scales with $\sqrt{t}$ thus eventually becomes much wider than $x-x'$ close to the peak. In this limit, $x'$ is considered small and we can perform a Taylor expansion of the Green's function arriving at
\begin{equation}
    \delta(x,t) = G(x,t)\int_{-\infty}^\infty \delta_0(x')\mathrm{d}x' - \partial_xG(x,t)\int_{-\infty}^\infty x'\delta_0(x')\mathrm{d}x' + \frac{\partial^2_xG(x,t)}{2}\int_{-\infty}^\infty x'^2\delta_0(x')\mathrm{d}x' + ....
\end{equation}

The leading order term is proportional to $G$, hence the height of the peak scales with $t^{-0.5}$, which is the case for a simple initial condition such as a delta function. However, the odd case yields
\begin{equation}
    \int_{-\infty}^\infty \delta_0(x')\mathrm{d}x' = 0.
\end{equation}
Thus the zeroth order moment of the initial condition is zero and leading order term is the dipole moment of the initial condition which scales according to $\partial_xG(x,t)$.

Taking the derivative of the Green's function we arrive at 
\begin{equation}
    \partial_xG(x,t) = -\frac{x}{2Dt}G(x,t).
\end{equation}

To find the location of the peaks of the derivative of the Green's function, we take a further derivative, giving 
\begin{equation}
    \partial^2_xG(x,t) = -\left[1-\frac{x^2}{2Dt}\right]\frac{G(x,t)}{2Dt} = 0
\end{equation}
which has solution $x = \pm\,\sqrt{2Dt}$.

This allows us to evaluate $\partial_xG(x,t)$ at the peaks to arrive at 
\begin{equation}
    \partial_xG(x,t) = \frac{\pm\exp[-1/2]}{2Dt\sqrt{2\pi}}
\end{equation}
which scales with $t^{-1}$ as observed in Fig.~\ref{fig:propagation_dan_1d}(c,f).

\section{Appendix H.  Quartic excitation}

In the section, we show that the propagation, spreading and skewing of a perturbation is largely insensitive to small variations in the initial condition. In the main text, the results where obtained for an initial gaussian profile $\delta_0\exp(-(x/w_0)^2/2)$, see top row of Fig.~\ref{fig:propagation_gaussian} (same figures as Fig.~\ref{fig:propagation} of the main text). 
One can reproduce the exact same results with an initial quartic profile $\delta_0\exp(-(x/w_0)^4/4)$, see bottom row of Fig.~\ref{fig:propagation_gaussian}. After a very short transient time, the profile is indeed indistinguishable from the initial gaussian case. 

\begin{figure}
    \centering
    \includegraphics[width=0.5\linewidth]{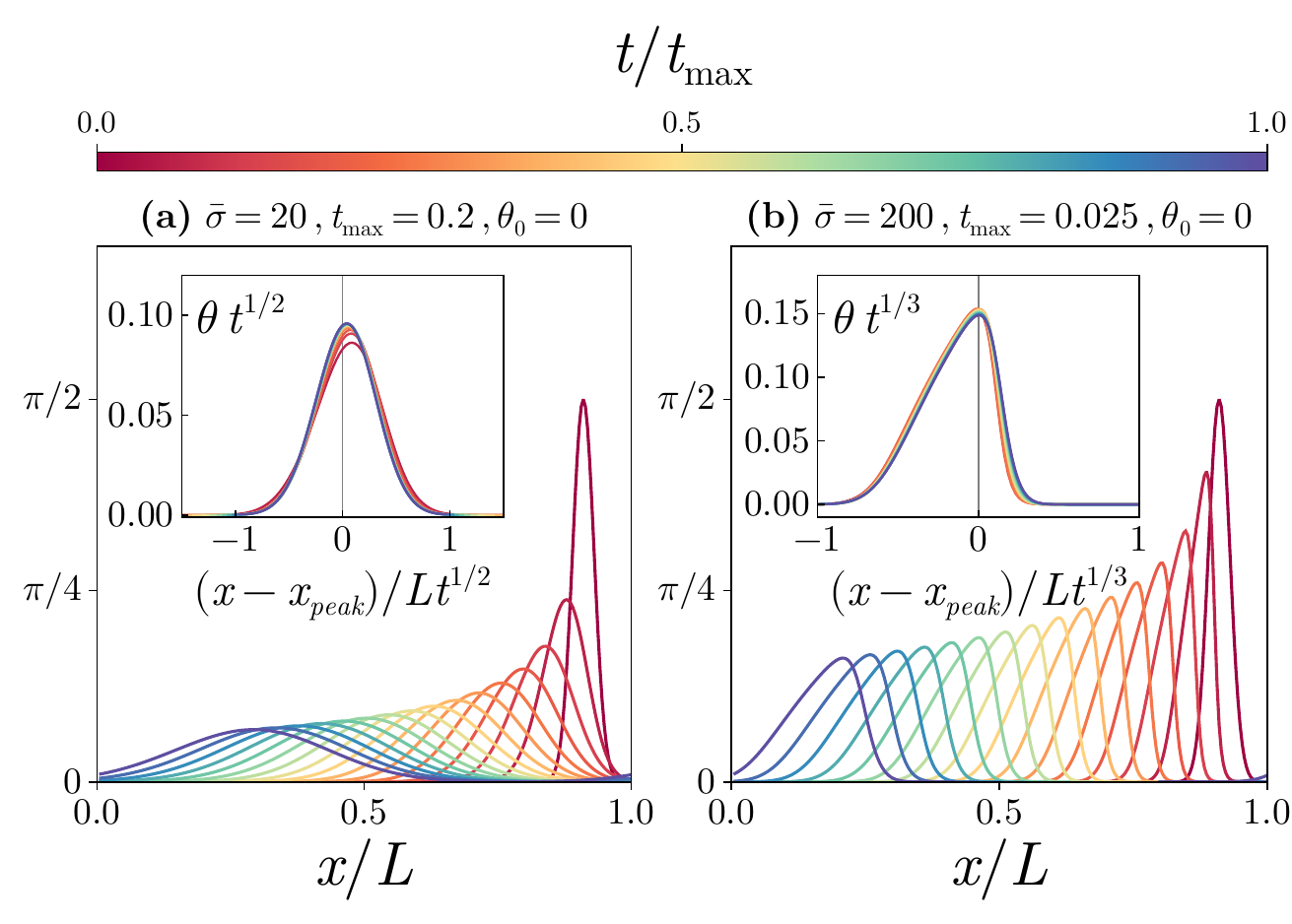}
    \includegraphics[width=0.5\linewidth]{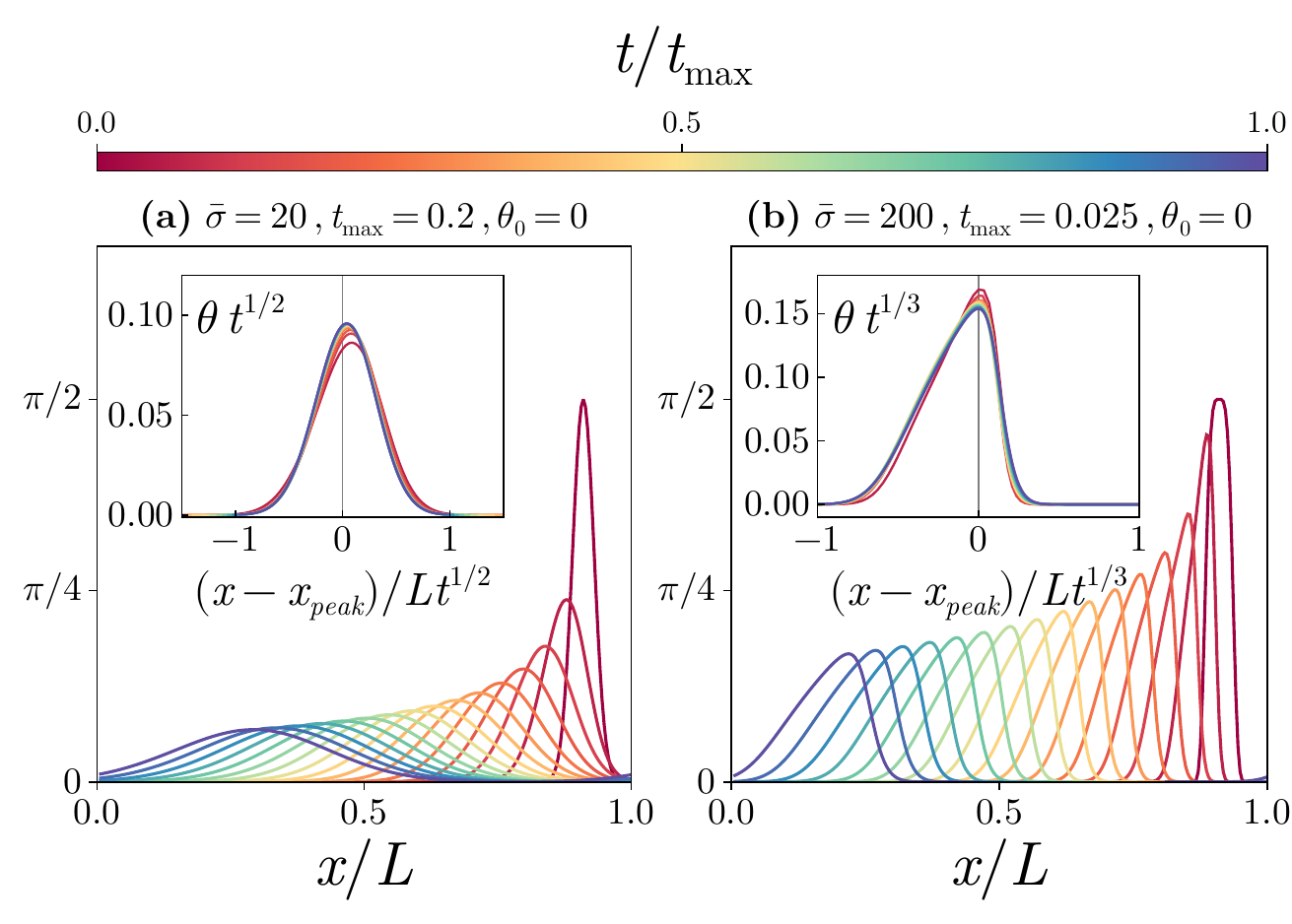}
   \caption{Time evolution of the profiles for perturbations initially \textbf{(top row)}  gaussian \textbf{(bottom row)}  quartic .
    \textbf{(a)} Propagation of a perturbation $\theta=\delta(x)$ across the system over time. Parameters: small non-reciprocity $\bar{\sigma} = 20, t_{\text{max}} = 0.2, N_{\text{mesh}} = 256, \alpha = 100, \theta_0=0$. \underline{Inset}: rescaled data $\theta\sqrt{t}$ as a function of $(x-x_{\text{peak}})/\sqrt{t} $.  
    \textbf{(b)} Same but for $\bar{\sigma} = 200, t_{\text{max}} = 0.025$. \underline{Inset}: rescaled data $\theta \,t^{1/3}$ as a function of $(x-x_{\text{peak}})/t^{1/3} $.}
    \label{fig:propagation_gaussian}
\end{figure}

\section{Appendix I. Derivation of the stationary solution of the TPP} \label{sec:derivation}

We start from Eq.~(\ref{eq:main_eq_1dof_1d}) describing the evolution of a $1d$ system:
\begin{equation}
\gamma \partial_t \theta
= K \partial_{xx} \theta + \bar{\sigma}\, \cos(\theta)\,\partial_x \theta 
\label{eq:SM_first_eq_derivation}
\end{equation}
A stationary profile $\theta(x)$ satisfies
\begin{equation}
0 = \partial_{xx} \theta(x) + \frac{\bar{\sigma}}{K}\, \cos\!\big(\theta(x)\big)\,\partial_x \theta(x)
\end{equation}
Thanks to the change of variable $p(x)=\partial_x \theta(x)$, we will reduce the order of our equation. Note that the chain rule implies
\begin{equation}
    \partial_{xx} \theta = \partial_xp = \frac{\partial p}{\partial\theta}\frac{\partial\theta}{\partial x}\,= \frac{\partial p}{\partial\theta}\,p
\end{equation}
Assuming a monotone profile ($p= \partial_x \theta\neq 0$), we can divide by $p$ and integrate: 
\begin{equation}
\frac{dp}{d\theta} = -\,\frac{\bar{\sigma}}{K}\, \cos \theta
\quad\Rightarrow\quad
p(\theta) = C - \frac{\bar{\sigma}}{K}\, \sin \theta
\end{equation}
One thus has
\begin{equation}
\partial_x \theta_s = C - \bar{\sigma}\, \sin \theta_s
\label{eq:SM_dtheta}
\end{equation}
where $C$ is a constant of integration that we will specify below. $C$ has units of inverse length. We now want to integrate this expression once more: we will separate the variables $\theta$ and $x$:
\begin{equation}
dx = \frac{d\theta}{C-\frac{\bar{\sigma}}{K}\, \sin \theta}
\qquad \text{such that } \qquad
\int dx = x-x_0 = \int \frac{d\theta}{C-\frac{\bar{\sigma}}{K}\, \sin \theta}
\label{eq:SMeq1}
\end{equation}
where $x_0$ is a constant of integration {that depends on the phase of the initial condition, $\theta_0$ in Eq.~(\ref{eq:spinwave})}. To perform the integration on the right hand side of Eq.~(\ref{eq:SMeq1}), we resort to the change of variable 
\begin{equation}
    t=\tan\left(\frac{\theta}{2}\right)
\end{equation}
which implies $\theta = 2\arctan t$ . Recall that 
    \begin{align}
        &\sin(2x)=2\sin x \, \cos x \\
        &\cos \arctan x = 1 / \sqrt{1+x^2} \\
        &\sin \arctan x = x / \sqrt{1+x^2} \\
        &d\arctan(x)/dx = 1 / (1+x^2)
    \end{align}
Following, one has 
\begin{equation}
    \sin \theta = \frac{2t}{1+t^2} \qquad \text{and} \qquad d\theta = \frac{2\,dt}{1+t^2}
\end{equation}
Thus
\begin{equation}
\int \frac{d\theta}{C-\bar{\sigma}\, \sin \theta/K}
= \int \frac{2\,dt}{\,C(1+t^2)-2\bar{\sigma}\, t/K\,}.
\end{equation}
Let us focus on the denominator and complete the square to make the next integration easier: 
\begin{equation}
    C(1+t^2)-2\bar{\sigma}\,t/K
= C\Big[(t-\tfrac{\bar{\sigma}}{KC})^2 + (\tfrac{\Delta}{C})^2\Big]
\end{equation}
where we defined $\Delta=\sqrt{C^2-\bar{\sigma}^2}$ .  Rewriting, one has
\begin{equation}
\frac{2}{C}\int \frac{dt}{(t-\tfrac{\bar{\sigma}}{KC})^2 + (\tfrac{\Delta}{C})^2} = \frac{2}{C}\int \frac{d(t - \bar{\sigma} /K C)}{(t-\tfrac{\bar{\sigma}}{KC})^2 + (\tfrac{\Delta}{C})^2}
\end{equation}
We now use the formula 
\begin{equation}
\int \frac{d u}{u^2+a^2}=\frac{1}{a} \arctan \frac{u}{a}
\end{equation}
to obtain 
\begin{equation}
\int \frac{d\theta}{C-\bar{\sigma}\, \sin \theta/K}
= \frac{2}{\Delta}\,
\arctan\!\Big(\frac{C\,t-\bar{\sigma}/K}{\Delta}\Big)
\end{equation} 
Now we can determine the value of the constant of integration $C$ by turning to determinate integrals in Eq.~(\ref{eq:SMeq1}). The integral over $x$ is simple:  $\int\limits_0^L dx = L$. 
Let us now consider the integral over $\theta$. 
When $\theta$ runs from 0 to $2\pi$, $t=\tan(\theta/2)$ runs from $-\infty$ to $+\infty$ and the arctan changes by $\pi$. 
One thus has
\begin{equation}
   \left.
   \frac{2}{\Delta}\,
\arctan\!\Big(\frac{C\,t-\bar{\sigma}}{\Delta}\Big) 
\right|_{\theta=0}^{\theta=2\pi} = \frac{2}{\Delta}\pi 
\end{equation}
Thus, because both integrals are equated,  and 
\begin{equation}
\Delta = \frac{2\pi}{L} \quad \text{and} \quad C = \sqrt{\left(\frac{\bar{\sigma}}{K}\right)^2 + \left(\frac{2\pi}{L}\right)^2  }
\end{equation}
Note that $\Delta$ and $C$ have units of 1/length. Finally, we isolate $t$. 
\begin{equation}
t = \frac{1}{C}\, \left[ \frac{\bar{\sigma}}{K} + \Delta \tan \left( \frac{\Delta}{2}(x-x_0)\right)\right]
\end{equation}
And $\theta = 2\arctan t$. In order to fit the data of Fig.~\ref{fig:concentration1}(a,b), we use $x_0 = 0.17$. 
\newline 

Finally, one can derive the width $w$ defined as 
\begin{equation}
    w=\frac{2\pi}{ \partial_x\theta(x=3L/4 )} \ .
\end{equation}
One could derive the expression obtained for $\theta$, but it is much simpler to use the expression Eq.~(\ref{eq:SM_dtheta}). Since $x=3L/4 $ is a fixed point of the profile,  $\theta(x=3L/4 ) = 3\pi/2$ at all times, for all $\bar \sigma$. Therefore, $\sin \theta(x=3L/4 )=-1$ and $\partial_x \theta = C + \bar \sigma$. Therefore, 
\begin{equation}
    w = \frac{2\pi}{(C + \bar \sigma/K)} = \frac{2\pi}{\sqrt{\left(\frac{\bar{\sigma}}{K}\right)^2 + \left(\frac{2\pi}{L}\right)^2}} \ .
\end{equation}
{Finally, we substitute in for the length scale $\lambda_{\bar{\sigma}} = K/\bar{\sigma}$ to arrive at}
\begin{equation}
    w = \frac{2\pi\,\lambda_{\bar{\sigma}}}{1+\sqrt{1 +\left(\frac{2\pi}{L}\lambda_{\bar{\sigma}} \right)^2 }}
\end{equation}

Let us derive the limits of $w$ for small and large $\bar \sigma$. Because $\lambda_{\bar{\sigma}} = K/\bar{\sigma}$,  $\bar \sigma \to 0$ implies $\lambda_{\bar{\sigma}} \to \infty$ and $\bar \sigma \to \infty$ implies $\lambda_{\bar{\sigma}} \to 0$. Following, one has 
\begin{equation}
     w \xrightarrow{\bar \sigma\, \to\, 0} L 
\end{equation}
and (recall that $\lambda_{\alpha} = \sqrt{K/\alpha}$)
\begin{equation}
     w 
     \xrightarrow{\bar \sigma\, \to\, \infty} 
     \pi \lambda_{\bar{\sigma}} 
     = \pi \lambda_{\alpha} \frac{\lambda_{\bar{\sigma}}}{\lambda_{\alpha}}
      = \pi \lambda_{\alpha} \frac{\bar \sigma}{\sqrt{K\alpha}}
\end{equation}
As such, the break down criteria $w = \lambda_{\alpha}$ occurs for $\frac{\bar \sigma}{\sqrt{K\alpha}} = 1/\pi \approx 0.318$ , cf. vertical grey line in Fig.~\ref{fig:concentration1}(d). 
\end{document}